\documentclass[twocolumn]{aastex63}

\usepackage{amsmath}
\usepackage{color}

\def\la{\mathrel{\hbox{\rlap{\hbox{\lower4pt\hbox{$\sim$}}}\hbox{$<$}}}}

\def\ga{\mathrel{\hbox{\rlap{\hbox{\lower4pt\hbox{$\sim$}}}\hbox{$>$}}}}

\usepackage{journal_names}

\usepackage{url}
\usepackage[utf8x]{inputenc}
\newcommand{\fmg}{\mbox{\ensuremath{\;\mathrm{.}\!\!^\textrm{m}}}}

\usepackage{graphicx,times}
\newcommand{\galfit}{\textsc{Galfit}}
\newcommand{\iraf}{\textsc{Iraf}}
\newcommand{\stsdas}{\textsc{Stsdas}}

\newcommand{\ellipse}{\textsc{Ellipse}}
\newcommand{\psf}{\textsc{PSF}}

\newcommand{\matplotlib}{\textsc{MATPLOTLIB}} 
\newcommand{\scipy}{\textsc{SCIPY}}
\newcommand{\numpy}{\textsc{NUMPY}} 
\newcommand{\pgplot}{\textsc{PGPLOT}}

\newcommand{\fref}{Figure~\ref}
\newcommand{\tref}{Table~\ref}
\newcommand{\Sref}{Section~\ref}

\usepackage{float}

\shorttitle{Distance and Peculiar Velocity of Norma Cluster}
\shortauthors{T. Mutabazi}

\begin{document}

\title{The Distance and Peculiar Velocity of the Norma cluster (ACO\,3627) 
using the Near-Infrared $\pmb{J}$ and $\pmb{K_s}$-band Fundamental Plane Relations.}

\correspondingauthor{T. Mutabazi}
\email{tmutabazi@must.ac.ug}

\author[0000-0002-9949-0403]{T. Mutabazi}
\affiliation{Department of Physics, Mbarara University of Science and Technology, P.O. Box 1410, Mbarara, Uganda}
\affiliation{Astrophysics, Cosmology and Gravity Centre (ACGC), Astronomy Department, 
University of Cape Town, Private Bag X3, Rondebosch, 7701, South Africa}



\begin{abstract}
We report distance measurements for the Norma cluster based on the near-infrared $J$- and $K_s$-band Fundamental Plane (FP) relations. Our simultaneous $J$ and $K_s$-band photometry analyses were performed using 31 early-type galaxies in the nearby Norma cluster obtained using the 1.4\,m InfraRed Survey Facility (IRSF) at the South African Astronomical Observatory. 
Our final $K_s$-band FP sample consists of 41 early-type galaxies from the Norma cluster observed using the IRSF and the New Technology Telescope (NTT) at the European Southern Observatory. This is the largest cluster sample used for peculiar velocity studies in the Great Attractor region to date. 
From the $K_s$-band FP, we find a distance to the Norma cluster of 4915\,$\pm$\,121\;km\;s$^{-1}$. The implied peculiar velocity for Norma is $44$\,$\pm$\,$151$\;km\;s$^{-1}$ which further supports a small peculiar velocity for the Norma cluster. 
\end{abstract}


\section{Introduction} \label{sec_intro}

While the Norma cluster (ACO\,3627) was identified and catalogued by \citet{Abell_89}, this cluster's massive nature and dynamical role in the Great Attractor (GA) region were not realised until the mid 1990s \citep{Kraan_96,Boehringer_96, Tamura_98}. The Norma cluster has since been identified by \citet{Woudt_08} as the richest cluster in the GA region. Studies in this region are affected by high levels of Galactic extinction and star-crowding due to its location at low Galactic latitudes. Some large scale structures are therefore expected to still remain hidden within this zone of obscuration \citep{Sorce_17, Courtois_19a} as evidenced by the recent discovery of the Vela supercluster \citep{Kraan_2017a}. Currently, there is significant improvement in our understanding of the large scale structures residing in the zone of avoidance through observations at near-infrared wavelengths where the effect of Galactic extinction significantly reduces \citep[e.g.][]{Macri_19}. Using telescopes with superior resolution such as the IRSF and NTT minimises the effects of high stellar density (star-crowding) in this region \citep[see e.g.][]{Said_016}. In addition, 21-cm radio observations \citep{Said_2016, Staveley_16, Kraan_18, Schroder_19} have revealed the abundance and massive nature of galaxies in this region. Such {\scshape{Hi}} observations are not affected by Galactic extinction but may suffer from radio contamination from neighbouring radio sources.

The Local Group has a well known peculiar velocity ($v_{_{\text{LG}}} = 627 \pm 22$~km\;s$^{-1}$) with respect to the CMB rest frame  in the direction $\ell = 276^\circ \pm 3^\circ$, $b = 30^\circ \pm 3^\circ$ \citep{Kogut_93} --- which is in the direction of Shapley \citep{Shapley_30, Raychaudhury_89, Scaramella_89} and Vela \citep{Kraan_2017a} superclusters. %
Despite more than two decades of study, the fractional contribution of the GA and Shapley supercluster on the Local Group's peculiar velocity 
remains a topic of debate \citep[see e.g. ][]{Smith_00, Lucey_05, Erdogdu_06, Kocevski_06, Lavaux_10, Carrick_15}. 

In the analysis by \citet{Lavaux_10}, they recovered the majority of the amplitude of the Local Group's peculiar velocity within 120$h^{-1}$\;Mpc but the direction of motion did not agree with the expected direction of the CMB dipole --- they suggested convergence may lie as far as 200$h^{-1}$\;Mpc which is supported by recent measurements using the 2M++ sample \citep{Carrick_15}.
While some authors found small contributions from the Shapley Supercluster (e.g., \citealt{Bardelli_00, Bolejko_08}), significant contributions from distant structures have been supported by e.g. \citet{Kocevski_06, Springob_14, Springob_16, Hoffman_17, Qin_2018, Tully_19, Said_20}. %

Although the zone of avoidance or zone of obscuration has received less attention, structures hidden within or behind this region may play a role in the general flow and may significantly contribute to the Local Group's motion. 
An example is the recent discovery of an extended supercluster of galaxies at $cz \sim 18 000$\;km\;s$^{-1}$, located at Galactic longitudes and latitudes $(\ell, b) \sim (272^\circ, 0^\circ)$ \citep{Kraan_2017a}. This previously hidden Vela supercluster was later kinematically confirmed by \citet{Courtois_19a}. %
Using Cosmicflows-3 which is a compilation of distances for $\sim$\,18000 galaxies \citep{Tully_16}, \citet{Courtois_17} found dominant features (repulsive and attractive), the major basin of attraction corresponding to an extension of the Shapley supercluster. This is in agreement with the previous findings based on the 6dF Galaxy Peculiar Velocity Survey %
where positive line-of-sight peculiar velocities were estimated, suggesting substantial mass overdensities originating at distances further away than the Great Attractor \citep{Springob_16}. %
There is currently increasing support for flow towards more distant structures such as Vela and Shapley superclusters \citep[see e.g. ][]{Kocevski_06, Springob_14, Staveley_16, Hoffman_17, Tully_19}. %

Located at \mbox{($\ell$, $b$, $v_{\rm{hel}}$)\,$=$\,(325$^\circ$, --7$^\circ$, \,$4871\pm 54$\,km\;s$^{-1}$)} --- see for example, \citet{Woudt_08}, the Norma cluster lies close to the anticipated centre of the GA \citep{Kolatt_95}. As such, a reliable estimate of the redshift-independent distance (and hence the peculiar velocity) of the Norma cluster is crucial to understanding the GA flow. Our previous measurements \citep{Mutabazi_14} show the Norma cluster to have a small peculiar velocity, consistent with zero (i.e.\ $v_{_{\rm{pec}}}$\,$=$\,$-$\,72\,$\pm$\,170\,km\;s$^{-1}$).

In order to improve the Fundamental Plane (FP) distance of Norma, we have made new $J$ and $K_s$-band photometric measurements using near-infrared (NIR) images from the 1.4\,m Japanese InfraRed Survey Facility (IRSF) at the Sutherland site of the South African Astronomical Observatory. We improve on our previous results by studying the effect of Galactic extinction using the $J-K_s$ colour for these new data. In addition, we have extended/improved our previous $K_s$-band FP analysis from $N$\,$=$\,31 to a total sample of 41 Norma cluster early-type galaxies (ETGs) selected from within $\frac{2}{3}R_{\rm A}$ where $R_{\rm A}$ is the Norma cluster's Abell radius.

In this paper, we therefore present (1) distance and peculiar velocity for the Norma cluster using the $J$ and $K_s$-band FP \citep{Djorgovski_87, Dressler_87_fp} analysis for 31 ETGs whose NIR images were obtained using IRSF (2) a combined $K_s$-band FP analysis for a total of 41 ETGs whose NIR images were obtained using the IRSF and NTT telescopes \citep[see][for telescope details]{Mutabazi_14}. For ETGs that were observed using both the IRSF and NTT, the NTT data was given priority due to its superior resolution and good seeing conditions. %
Note however that both datasets have high quality, highly resolved data (NTT and IRSF have pixel scales of 0.29 and 0.45 arcsec per pixel, respectively) which are very well suited for this study.

This paper is structured as follows: Sample selection and observations for our FP sample is discussed in $\S$\;\!\ref{sample_data} while the photometric data and analysis are presented and discussed in $\S$\;\!\ref{data_analysis}. A re-calibration of the NIR reddening maps is presented in $\S$\;\!\ref{extinc_recalib}. %
In $\S$\;\!\ref{offset}, we present the measured zero-point offset using the FP relation. %
The distance and peculiar velocity measurements are presented in $\S$\;\!\ref{dist_vpec_fp}. The discussion and conclusions of our findings are presented under $\S$\;\!\ref{sec_discuss}.

%
Unless stated, we have adopted standard cosmology with $\Omega_\text{m}$\,$=$\,$0.27$, \mbox{$\Omega_{\Lambda}$\,$=$\,$0.73$} and \mbox{$\text H_0$\,$=$\,$70.5\; \rm{km} \; \rm s^{-1} \!\;\rm{Mpc}^{-1}$} \citep{Hinshaw_09}. While measuring the distance of the Norma cluster relative to the Coma cluster, we used \mbox{$z_{_{\rm{CMB}}} = 0.02400 \pm 0.00016$} for the Coma cluster \citep{Hudson_04}. 

\section{Observations and data reduction} \label{sample_data} 

\subsection{Sample selection}

The ETGs used in the Norma cluster sample were such that (1) 2096\,km\;s$^{-1}$~$<v_{\rm hel}<$~7646\,km\;s$^{-1}$ where $v_{\rm hel}$ is the 
galaxy's heliocentric velocity (2) lie within $\frac{2}{3}$ of the Abell radius, 
and (3) we had successfully measured the central velocity dispersion using the 2dF spectrograph. 
The above selection criteria resulted in a total of 31 ETGs with $J$ and $K_s$-band images from the IRSF telescope. Ten of these 31 ETGs with IRSF images are unique to the IRSF while 21 ETGs have NIR images from both the IRSF and NTT.

For the Coma cluster sample, galaxies selected were those typed as E or E/S0 or S0 by \citet{Dressler_80}, were identified and confirmed as Coma cluster members based on their redshifts, and had a central velocity dispersion in the Sloan Digital Sky Survey  Data Release~8 \citep{Aihara_11}. In addition, only those which we successfully measured the FP photometry parameters from the 2MASS Atlas images were selected. This resulted in a Coma cluster sample of 121 ETGs.

\subsection{Observations with the IRSF}
 
The IRSF telescope is equipped with SIRIUS, a three-colour band camera with three 1024$\times$1024 pixels HgCdTe infrared detectors and is capable of simultaneous imaging in the $J$, $H$ and $K_s$ passbands \citep{Nagashima_99}. %
Our IRSF observations were carried out using a total integration time of 600\,s split over 20 short exposures of 30\,s each. Standard NIR data reduction procedures (such as dark subtraction, flat-fielding, sky subtraction) were applied before combining the 20 dither frames into one science image. Note that both the NTT and IRSF telescopes are suitable for the high stellar density GA region since they are equipped 
with the SOFI and SIRIUS imaging instruments which have a low pixel scale, that is, 0.29\;arcsec per pixel and 0.45\;arcsec per pixel, respectively. It is also important to note that the combined $K_s$-band data (IRSF+NTT) is the largest cluster sample ($N$\,$=$\,$41$) used for peculiar velocity studies in the GA region to date.

To ensure that all our photometry data sets are on the same photometric system, the astrometric and photometric calibrations were performed using the 2MASS Point Source Catalogue \citep{Skrutskie_06}. During the photometric calibration process, we applied a small correction due to the slight difference between the 2MASS, NTT and IRSF filters (see e.g., \citealt{Carpenter_01}). %

\section{Photometry data analysis} \label{data_analysis}

In order to apply the FP relation, measurements of the effective radii and mean effective surface brightnesses are required. These were measured through galaxy surface brightness profile fitting.
 
\subsection{Galaxy surface brightness profile fitting} \label{ellipses}

\begin{figure}[htb!]
 \centering 
   \begin{tabular}{c} 
     \includegraphics[width=0.45\textwidth]{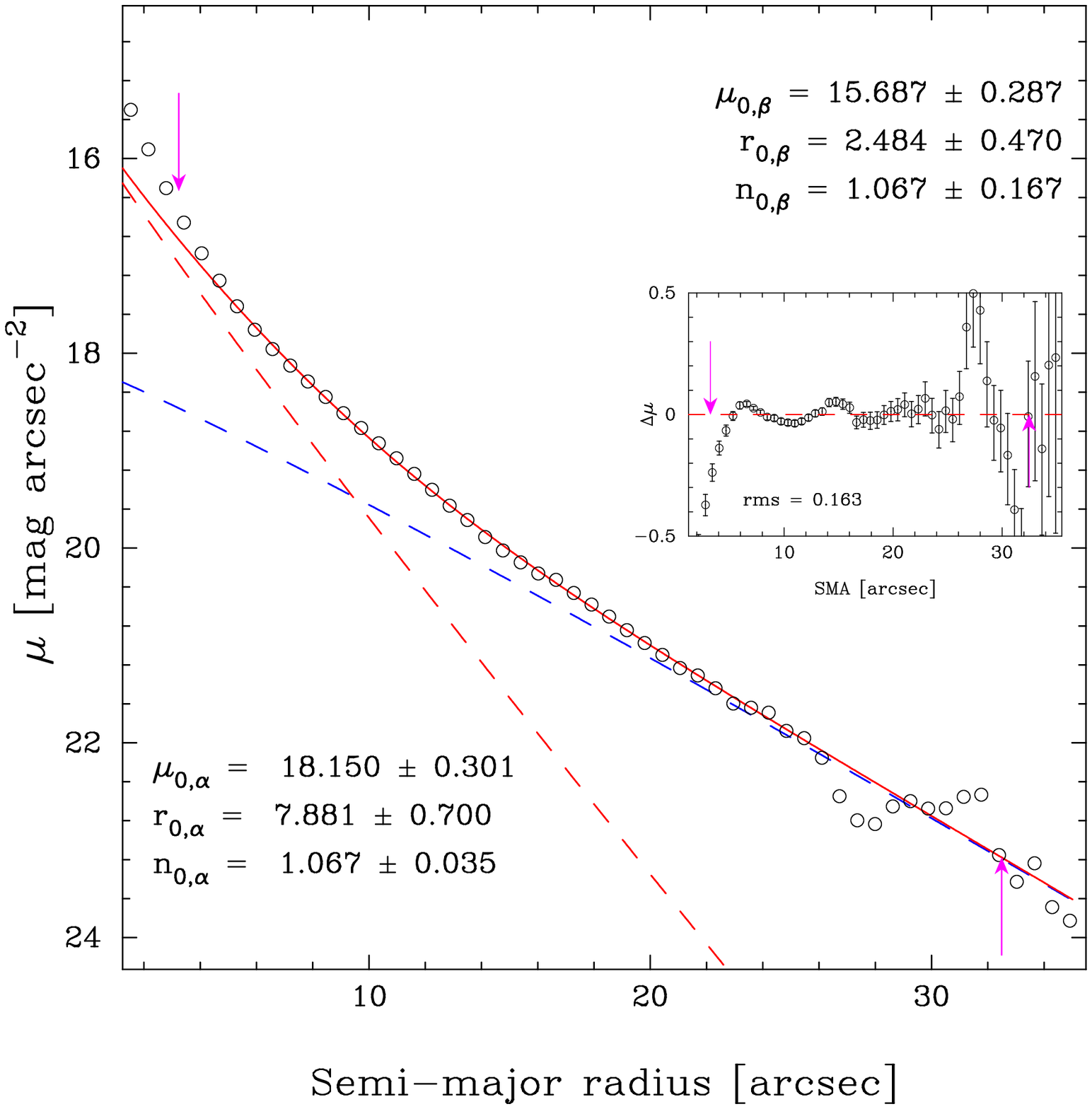} \\
%
~\\~\\
     \includegraphics[width=0.45\textwidth]{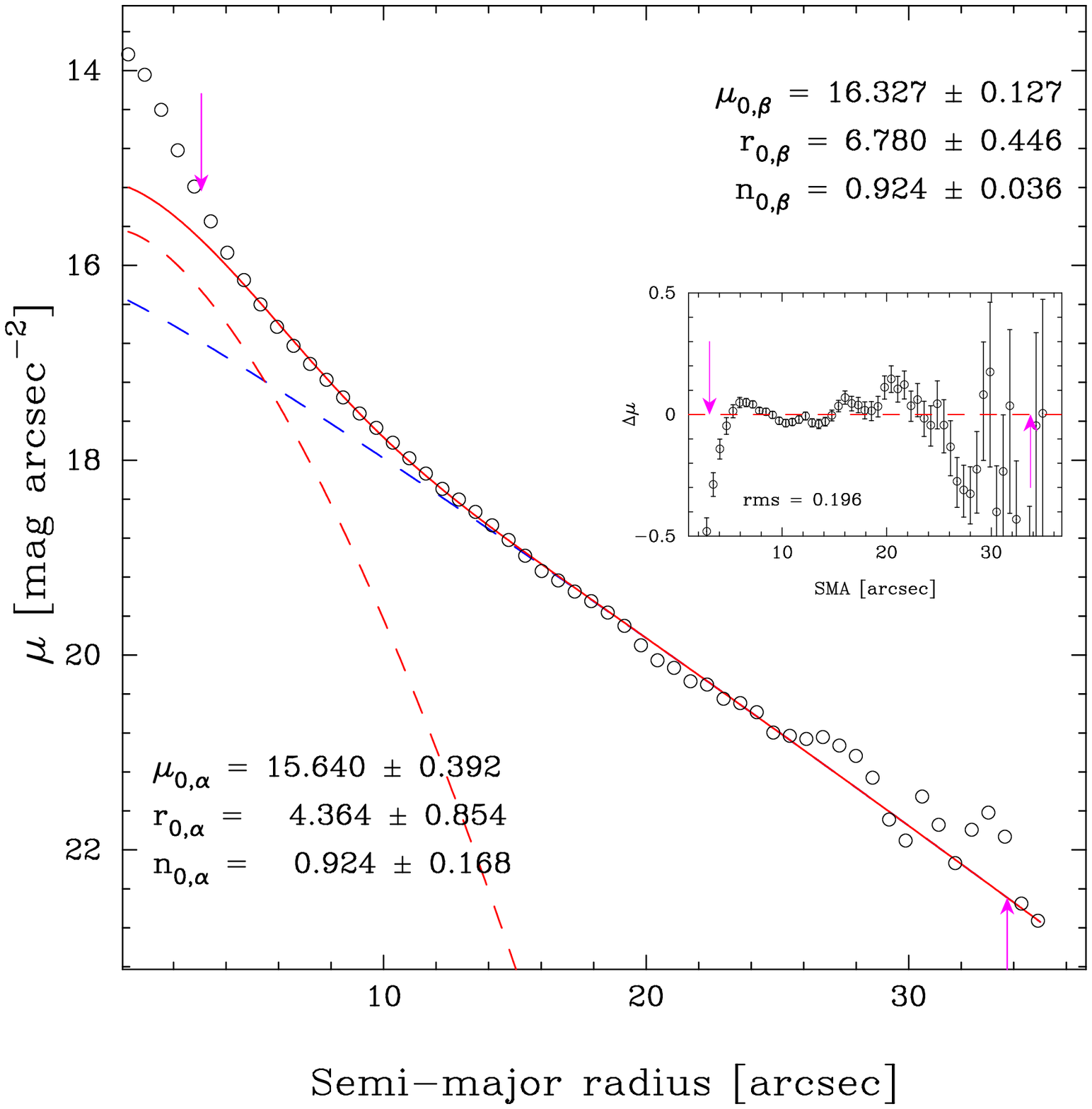}
  \end{tabular}
\caption{\label{extrapolate:singDsersics_Nfit1d} Double S\'{e}rsic component fits to the galaxy surface brightness profile for WKK\,6250 in the $J$ (top panel) and $K_s$-band (bottom panel). The red and blue dashed lines represent the individual S\'{e}rsic components while the red solid line represents the combined fit (best fit to the galaxy surface brightness profile). The small magenta arrows represent twice the seeing (FWHM) and the radius where the galaxy surface brightness reaches the sky background noise level. Only the data within the range indicated by these two magenta arrows were used to fit the galaxy surface brightness profile. 
}
\end{figure}

For accurate photometry, stars in each galaxy field were subtracted after which a reliable estimate of the sky background was calculated using an annulus as described in \citet{Mutabazi_14}. We performed the photometric analysis (for both Norma and Coma galaxies) using the \ellipse\ \citep{Jedrzejewski_87} task in \iraf. %
The resulting surface brightness profiles were fitted using a combination of two S\'{e}rsic functions. The best fit was extrapolated to infinity so as to estimate the total extrapolated magnitude. The effective radius was then measured through interpolation. To correct for the seeing (\psf) effects, the \galfit\ algorithm \citep{Peng_10} was used with and without a \psf\ convolution. 
Figure~\ref{extrapolate:singDsersics_Nfit1d} is an example of the fitted profiles for WKK\,6250, one of the Norma cluster's galaxies from the IRSF data. The top panel is the $J$-band while the bottom panel represents the $K_s$-band. 
The fit data were restricted to within the radius ranging from twice the seeing to where the galaxy flux averages to 1$\sigma_s$ above the sky background. 
These are represented by the two small magenta arrows. The black solid line is the best fit which is a combination of the two S\'{e}rsic component fits represented by the red and blue dashed lines. The inset shows the residuals. %

\subsection{Photometry comparisons} \label{systematics}

\begin{figure}[htb!]
\centering
\includegraphics[width=0.43\textwidth]{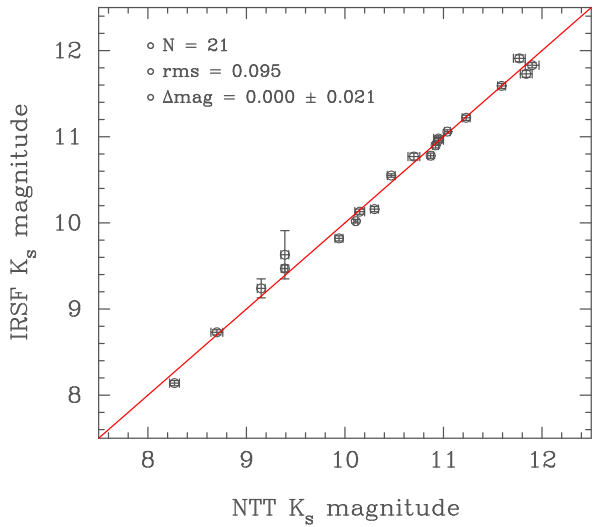}
\caption{\label{phot-comp-ntt-irsf}Comparison between the total extrapolated magnitudes for the 21 ETGs observed using both the NTT and IRSF telescopes. The red solid line shows a one-to-one relation. Both the average magnitude and rms were computed based on this line.}
\end{figure}

As a photometric reliability check, we took advantage of the high quality, high resolution images provided by both the NTT (pixel scale of 0.29 arcsec per pixel) and IRSF (pixel scale of 0.45 arcsec per pixel) telescopes. We compared the $K_s$-band total extrapolated magnitudes for an overlapping sample of 21 ETGs (ETGs which are common to both the IRSF and NTT data). The comparison shows excellent agreement as seen in Figure \ref{phot-comp-ntt-irsf}, i.e., an average difference in total extrapolated magnitudes of $0.000\pm 0.021$\,mag although with a relatively large scatter of 0.095 mag. %
This scatter is dominated by WKK\,6204 with a very uncertain IRSF $K_s$-band magnitude. When this galaxy (WKK\,6204) is excluded from the comparison between the IRSF and NTT magnitudes, the mean difference is $0.0125\pm 0.0180$ mag and the rms reduces from 0.095 to 0.080. However, the effect on the IRSF Ks-band Fundamental Plane analysis is not very significant -- the rms from the FP fit decreases slightly from 0.0778 to 0.0776 when WKK\,6204 is included and excluded, respectively. The source of the large uncertainty on the IRSF $K_s$-band magnitude for this galaxy (WKK\,6204) is not very clear since the error/uncertainty in the IRSF $J$-band magnitude is significantly lower than the IRSF $K_s$-band magnitude for this same galaxy (0.07 mag in $J$ and 0.28 mag in $K_s$-band). It should however be noted that there were sky gradients in the observed field for this galaxy using IRSF as compared to NTT possibly due to a superior resolution and better seeing conditions associated with NTT. These sky gradients are most likely due to a nearby bright galaxy in the field of WKK\,6204. Therefore, the large uncertainty on the IRSF magnitude for WKK\,6204 could be due to the error associated with the measured sky background in the $K_s$-band as a result of this nearby galaxy. The effect may be more pronounced in the $K_s$ image since the nearby galaxy is brighter in the $K_s$ than in the $J$-band. %
Note that the central velocity dispersions used in our analysis were taken from the Sloan Digital Sky Survey Data Release 8 \citep{Aihara_11} for the Coma cluster sample. For the Norma cluster sample, data using the fibre spectroscopy on the 2dF facility \citep{Lewis_02} on the 3.9\,m Anglo-Australian Telescope were used.  
%
 
\subsection{Coma cluster sample -- Photometric analysis}

For the Coma cluster, 2MASS Atlas images were used in the photometric analysis. For photometric quality check, we have compared our measurements and the 2MASS~XSC for the 121 Coma ETGs. The mean differences \mbox{(2MASS -- This work)} between our measured total extrapolated magnitudes and the values taken from the 2MASS~XSC \citep{Skrutskie_06} are $-$\,0.003\,$\pm$\,0.014~mag and $-$\,0.015\,$\pm$\,0.009~mag in the $J$ and $K_s$-band, respectively. The photometric results and the corrected FP variables for the Coma sample ($N$\,$=$\,121) are presented in 
Table~\ref{Coma_JKs_phot}. %

\subsection{Galactic extinction: re-calibration using NIR colours} \label{extinc_recalib}

As Norma is at Galactic latitude, $b=-7$, the Galactic extinction is relatively large. Note that for the Coma cluster, the effect of Galactic extinction is very small, i.e., $\sim$\,0.007~mag and 0.003~mag in the $J$ and $K_s$-band, respectively. %
%
The \citet{Schlegel_98} NIR reddening maps have been found to over-estimate the Galactic extinction especially at low Galactic latitudes \citep{Bonifacio_00, Yasuda_07, Schroder_07, Schlafly_11}. A re-calibration of the NIR reddening maps based on the \mbox{$J-K_s$} colour for the Norma and Coma cluster ETGs used in our FP analysis is presented here. This closely follows the re-calibration by \citet{Schroder_07}. 

\begin{figure}[htb!]
 \centering
 \begin{tabular}{c}
\includegraphics[width=0.43\textwidth]{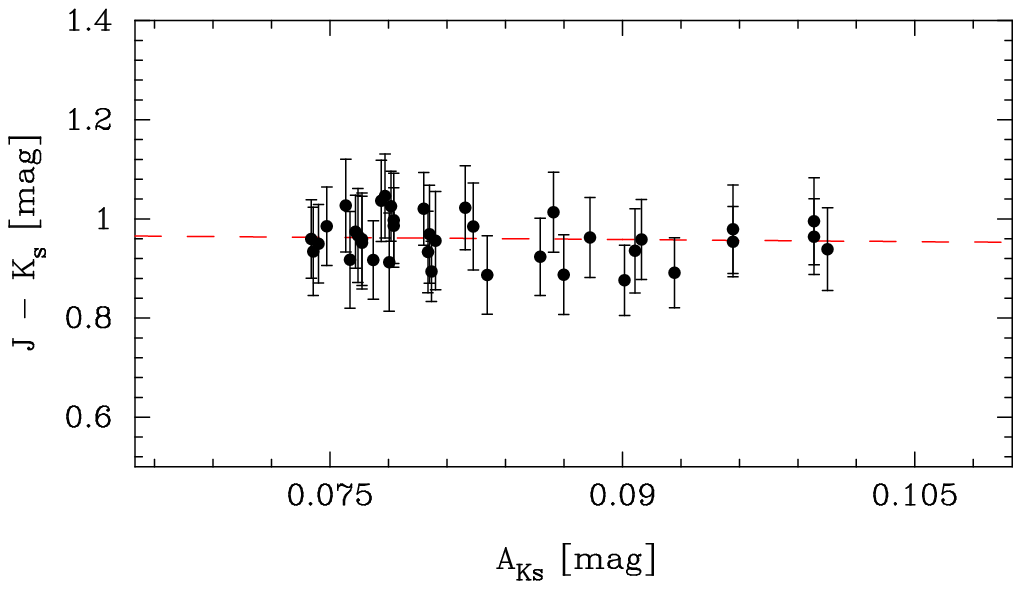} \\
~\\
\includegraphics[width=0.43\textwidth]{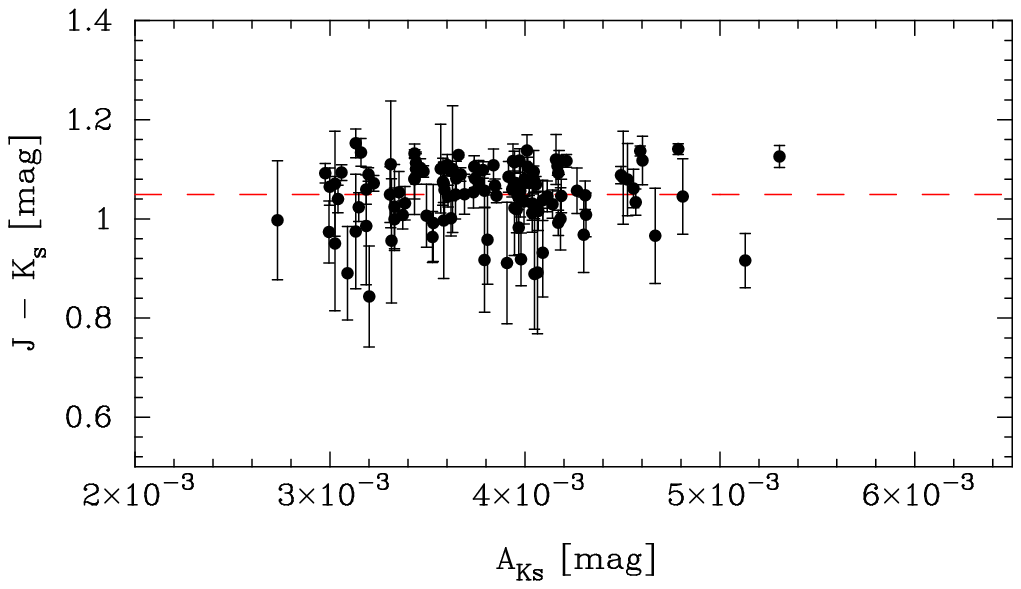}
\end{tabular}
\caption[$J$ and $K_s$-band photometry for the Coma sample]{\label{extinc_JKs}Measuring the extinction correction factor using the $(J-K_s)$ colour for the Norma (top panel) and the Coma (bottom panel) samples. The y-axis is the $(J-K_s)$ colour which has been corrected for both Galactic extinction and redshift effects. %
}
\end{figure}

The extinction in the $J$ and $K_s$ bands, assuming the \citet{Fitzpatrick_99} reddening law, with $R_{\rm v}=3.1$ is given by
\begin{equation} 
  A_J = 0.937 E(B-V), \qquad A_{K_s} = 0.382 E(B-V), \label{ext_eq2}
\end{equation}
where $E(B-V)$ is the colour excess or selective extinction. %
The redshift ($k$-correction) and extinction-corrected $(J-K_s)$ colour is given by
\begin{equation} 
   (J-K_s)^o = (J-K_s) - 0.555 \, E(B-V) + 4.0\,z; \label{extinc_factor1}
\end{equation}
where, $ (J-K_s)$ is the measured NIR colour with no corrections applied while $(J-K_s)^o$ is the NIR colour, corrected for both redshift ($k$-correction, see $\S$\;\!\ref{k_correct}: Equation~\ref{k-corr}) and Galactic extinction effects. The colour excess, $E(B-V)$ values, were obtained using the DIRBE/IRAS reddening maps \citep{Schlegel_98}. %
\fref{extinc_JKs} shows the least-squares fit to the Norma sample (top panel) and Coma sample (bottom panel). The $J$ and $K_s$ aperture magnitudes were measured using aperture radii of 5 arcsec. The best fit which is represented by the red dashed line is
{\small{
\begin{eqnarray}
\text{Norma:} ~ (J-K_s)^o = -0.283\pm0.087 A_{K_s} + 0.984\pm0.037, \label{eq1_norma}\\
\text{Coma:~} ~ (J-K_s)^o = -0.096\pm0.056 A_{K_s} + 1.050\pm0.067. \label{eq2_coma} 
\end{eqnarray}
}}
Equations~\ref{eq1_norma} and \ref{eq2_coma} can be expressed in the form
\begin{equation}
  (J-K_s)^o = aA_{K_s} + b \, ; \label{extinc_factor2}
\end{equation}
where $A_{K_s}$ is the extinction in the $K_s$-band which is given by Equation~\ref{ext_eq2}, $a$ and $b$ are fit parameters which were measured through simple least-squares regression. %

The reddening law for the DIRBE/IRAS reddening maps \citep{Schlegel_98} following the derivation in \citet{Schroder_07} can be written as:
\begin{equation}
J - K_s = \left(J - K_s\right)^o + \left[ \frac{E(J-K_s)}{A_{K_s}} \right] A_{K_s} \label{jtm_1}
\end{equation}
For the true extinction-corrected colour, $\left(\widetilde{J - K_s}\right)^o$, and true $K_s$-band extinction, $\tilde{A}_{K_s}$, we have
\begin{equation}
J - K_s = \left(\widetilde{J - K_s}\right)^o + \left[ \frac{E(J-K_s)}{A_{K_s}} \right] \tilde{A}_{K_s}.\label{jtm_2}
\end{equation}
Combining Equations~\ref{extinc_factor2} and \ref{jtm_1} yields
\begin{equation}
J - K_s = a A_{K_s} + b + \left[ \frac{E(J-K_s)}{A_{K_s}} \right] A_{K_s}. \label{jtm_3}
\end{equation}
Equating Equations~\ref{jtm_2} and \ref{jtm_3} leads to
\begin{equation}
\left(\widetilde{J - K_s}\right)^o = aA_{K_s} + b + \left( A_{K_s} - \tilde{A}_{K_s} \right) \left( \frac{E(J-K_s)}{A_{K_s}} \right)\cdot \label{jtm_4}
\end{equation}
The trend in Equations~\ref{eq1_norma} and \ref{eq2_coma} 
implies a multiplicative factor, $f$, between the applied and true extinction, that is,
\begin{equation}
\tilde{A}_{K_s} = f A_{K_s}. \label{jtm_5}
\end{equation}
Substituting Equation~\ref{jtm_5} into \ref{jtm_4} results in 
\begin{equation}
\left(\widetilde{J - K_s}\right)^o = a + \left[\left(1-f\right) \frac{E(J-K_s)}{A_{K_s}}\right] A_{K_s} + b. \label{jtm_6}
\end{equation}
Note that the true colour, $\left(\widetilde{J - K_s}\right)^o$, is independent of extinction, $A_{K_s}$. This implies that 
\begin{equation}
\left(1-f\right) \frac{E(J-K_s)}{A_{K_s}} = 0. \label{jtm_7}
\end{equation}
Equation~\ref{jtm_7} implies that
\begin{equation} 
    f = 1 + a \left[\frac{E(J-K_s)}{A_{K_s}} \right]^{-1}. \label{extinc_factor_final}
\end{equation}
From the measured slope (Equations~\ref{eq1_norma} and \ref{eq2_coma}) and the selective extinction given by Equation~
\ref{ext_eq2}, the extinction correction factors computed from Equation~\ref{extinc_factor_final} are \mbox{$0.81\pm0.06$} and \mbox{$0.93\pm0.05$} for the Norma and Coma samples, respectively. The challenge in our calibration is that, our sample covers a very small area on the sky and hence a very small range of extinction values (see the top and bottom panels of \fref{extinc_JKs}). It is however worth the effort given our high quality (high resolution) photometric data.  After applying these extinction correction factors (0.81 for Norma and 0.93 for Coma ETGs), the mean differences between the \citet{Schlafly_11} $K_s$-band extinction and our values are relatively small, that is, 0.0081 mag (rms\,=\,0.0008 mag) and 0.0003 mag (rms\,=\,0.000045 mag) for the Norma and Coma ETGs, respectively. The extinction correction factor of \mbox{$0.81\pm0.06$} implies a range of 0.75\,--\,0.87 which translates into a difference of $\sim$\,43\;km\;s$^{-1}$ in Norma's distance. 

\subsection{Corrected effective surface brightnesses} \label{k_correct}

The mean effective surface brightness ($\langle \mu_e \rangle$) was computed using:  
\begin{equation}
\langle \mu_e \rangle = m_{\rm{tot}} + 2.5 \log \left( 2\pi r^2_e \right) - A_{\lambda} + k_{\lambda} - 10 \log \left( 1+z \right),
\end{equation}
where $m_{\text{tot}}$ is the total extrapolated magnitude corrected for the star-subtraction and sky background effects. $A_{\lambda}$ is the Galactic extinction, $k_{\lambda}$ is the $k$-correction, and 
$10 \log \left( 1 + z \right)$ is a cosmological dimming correction term ($\lambda$ denotes the filter/band -- i.e., $J$, and $K_s$). The effective radii for both the Norma and Coma cluster ETGs were determined by measuring the radius of a circle containing half the total flux. These were thereafter corrected for seeing effects (measured using \galfit). The applied $k$-correction was adopted from \citet{Glass_99}, i.e., %
\begin{equation}
k_{J} = 0.7z, \qquad k_{K_s} = 3.3z. \label{k-corr}
\end{equation}
We have compared our adopted $k$-corrections with those of \citet{Chilingarian_10}. The mean differences (for the $K_s$-band) are 0.022 mag (rms\,=\,0.004 mag) and 0.039 mag (rms\,=\,0.005 mag) for the Norma and Coma samples, respectively. Using the \citeauthor{Chilingarian_10} $k$-correction values lowers the distance of the Norma cluster by $\sim$\,67\;km\;s$^{-1}$. %

Table~\ref{Norma_JKs_phot} shows the results for the Norma cluster sample where the final fully corrected FP variables ($r_e$, $\langle \mu_e \rangle$, and $\log \sigma$) are presented (also see Table~\ref{Coma_JKs_phot} for the Coma sample). %
Note that for the Norma cluster sample, $m_{\rm{tot}}$ was first corrected for star-subtraction effects (which are --\,0.0143 and {\mbox{--\,0.0135}} mag in the $J$ and $K_s$-band, respectively). An additional small correction (--\,0.011 and \mbox{--\,0.008} mag in the $J$ and $K_s$-band, respectively) due to sky background variation was also applied. This correction was measured (in our analysis) through simulations conducted using our IRSF $J$ and $K_s$-band images.

\section{Norma Distance: FP zero-point offset} \label{offset}

\subsection{Construction of the Fundamental Plane data sets} \label{fp_data_sets}

The FP relation 
requires the effective radius, $r_e$, mean effective surface brightness, $\langle \mu_e \rangle$, and the central velocity dispersion, $\sigma$, that is, 
\begin{equation}
\log r_e = a \log \sigma + b \langle\mu_e\rangle + c, \label{fp_eq}
\end{equation}
where $a$ and $b$ are the FP slopes and $c$ is the intercept. %

In \tref{Norma_JKs_phot}, we present the FP data for the Norma ETGs ($N$\,$=$\,31). 
Presented in this table are the central velocity dispersions and the photometric results obtained using the IRSF NIR $J$ and $K_s$-band images. The Galactic extinction correction applied to $\langle \mu_e \rangle$ is from the re-calibration presented in \Sref{extinc_recalib}, i.e., \mbox{$A_J = 0.937\,f\,E(B-V)$} and \mbox{$A_{K_s} = 0.382\,f\, E(B-V)$} where $f=0.81\pm0.06$ for the Norma cluster sample and $f=0.93\pm0.05$ for the Coma cluster calibration sample. 
\begin{table*}
\centering
\caption[Norma cluster $J$ and $K_s$ photometry results]{\label{Norma_JKs_phot} Norma cluster $J$ and $K_s$-band photometry results from f\mbox{}itting and extrapolating the galaxy surface brightness prof\mbox{}iles. NIR images taken using the IRSF telescope were used for the photometry analysis. The total extrapolated magnitudes presented here have been corrected for only, the star-subtraction and sky background effects. The $J$ and $K_s$-band mean ef\mbox{}fective surface brightness in $\rm{mag}\;\rm{arcsec}^{-2}$ have been corrected for the Galactic extinction, redshift, and the cosmological dimming ef\mbox{}fects. The Galactic extinction values are from the re-calibration presented under \Sref{extinc_recalib}, where, \mbox{$A_J = 0.937\,f\,E(B-V)$} and \mbox{$A_{K_s} = 0.382\,f\,E(B-V)$}. 
Both the aperture correction and run offset have been applied to the central velocity dispersion values (in dex) presented here. The FP variables, that is, effective radii, mean effective surface brightnesses and the central velocity dispersions presented here have been fully corrected. These were used in the final FP fitting and analysis.}
\begin{tabular}{c@{\hskip 0.15in}  r@{\hskip 0.15in}  r@{\hskip 0.15in}  r@{\hskip 0.15in}  r@{\hskip 0.15in}  c@{\hskip 0.15in}  c@{\hskip 0.15in}  c@{\hskip 0.15in}  c@{\hskip 0.15in}  c}  \\ \hline\hline
\multicolumn{1}{c@{\hskip 0.15in} }{Identification}  & \multicolumn{2}{c@{\hskip 0.15in} }{Total magnitude} &  \multicolumn{2}{c@{\hskip 0.15in} }{Effective radius} &  \multicolumn{2}{c@{\hskip 0.15in} }{Mean eff. surf. brightness} &  \multicolumn{2}{c@{\hskip 0.15in} }{Extinction} &  \multicolumn{1}{c}{$\log \sigma$} \\ 
 & \multicolumn{1}{c@{\hskip 0.15in} }{$J$}  & \multicolumn{1}{c@{\hskip 0.15in} }{$K_s$}  & \multicolumn{1}{c@{\hskip 0.15in} }{$J$} & \multicolumn{1}{c@{\hskip 0.15in} }{$K_s$} & \multicolumn{1}{c@{\hskip 0.15in} }{$J$} & \multicolumn{1}{c@{\hskip 0.15in} }{$K_s$}  & \multicolumn{1}{c@{\hskip 0.15in} }{$A_J$} & \multicolumn{1}{c@{\hskip 0.15in} }{$A_{K_s}$}  &         \\ 
\multicolumn{1}{c@{\hskip 0.15in} }{(1)} & \multicolumn{1}{c@{\hskip 0.15in} }{(2)}  & \multicolumn{1}{c@{\hskip 0.15in}}{(3)}  & \multicolumn{1}{c@{\hskip 0.15in} }{(4)} & \multicolumn{1}{c@{\hskip 0.15in} }{(5)} & \multicolumn{1}{c@{\hskip 0.15in} }{(6)} & \multicolumn{1}{c@{\hskip 0.15in} }{(7)}   & \multicolumn{1}{c@{\hskip 0.15in} }{(8)}  & \multicolumn{1}{c@{\hskip 0.15in} }{(9)} & \multicolumn{1}{c}{(10)}    \\ \hline
WKK6019	&		10.87$\pm$0.06	&9.82$\pm$0.03	&	4.56$\pm$0.50	&4.22$\pm$0.23	&	15.98$\pm$0.24	&14.81$\pm$0.12	&	0.172	&0.070	&2.409$\pm$0.010 \\
WKK6047	&		12.81$\pm$0.06	&11.83$\pm$0.03	&	3.07$\pm$0.22	&2.92$\pm$0.13	&	17.09$\pm$0.16	&16.04$\pm$0.10	&	0.173	&0.070	&2.010$\pm$0.015 \\
WKK6075	&		12.95$\pm$0.06	&11.92$\pm$0.03	&	2.32$\pm$0.27	&2.14$\pm$0.06	&	16.56$\pm$0.26	&15.38$\pm$0.07	&	0.181	&0.074	&2.099$\pm$0.017 \\
WKK6116	&		10.50$\pm$0.06	&9.47$\pm$0.04	&	7.48$\pm$1.00	&7.59$\pm$0.70	&	16.71$\pm$0.30	&15.71$\pm$0.21	&	0.163	&0.066	&2.344$\pm$0.012 \\
WKK6148	&		12.67$\pm$0.05	&11.69$\pm$0.03	&	3.84$\pm$0.18	&3.33$\pm$0.14	&	17.39$\pm$0.11	&16.13$\pm$0.10	&	0.155	&0.063	&1.998$\pm$0.019 \\
WKK6180	&		11.16$\pm$0.06	&10.13$\pm$0.03	&	6.84$\pm$0.69	&7.18$\pm$0.37	&	17.12$\pm$0.23	&16.24$\pm$0.12	&	0.159	&0.065	&2.308$\pm$0.011 \\
WKK6183	&		11.13$\pm$0.06	&10.16$\pm$0.03	&	5.32$\pm$0.82	&5.58$\pm$0.42	&	16.68$\pm$0.34	&15.66$\pm$0.17	&	0.155	&0.063	&2.377$\pm$0.012 \\
WKK6193	&		11.10$\pm$0.07	&10.00$\pm$0.03	&	6.48$\pm$0.68	&6.60$\pm$0.28	&	16.97$\pm$0.24	&15.99$\pm$0.10	&	0.159	&0.065	&2.226$\pm$0.010 \\
WKK6204	&		10.35$\pm$0.07	&9.63$\pm$0.28	&	6.87$\pm$1.12	&5.22$\pm$3.16	&	16.37$\pm$0.36	&14.98$\pm$1.34	&	0.160	&0.065	&2.499$\pm$0.009 \\
WKK6207	&		11.05$\pm$0.07	&9.90$\pm$0.03	&	4.38$\pm$0.83	&6.39$\pm$0.52	&	16.21$\pm$0.42	&15.78$\pm$0.18	&	0.152	&0.062	&2.241$\pm$0.012 \\
WKK6221	&		11.31$\pm$0.06	&10.77$\pm$0.04	&	9.14$\pm$1.05	&7.47$\pm$0.38	&	17.89$\pm$0.26	&16.62$\pm$0.11	&	0.164	&0.067	&2.038$\pm$0.017 \\
WKK6229	&		12.56$\pm$0.07	&11.59$\pm$0.03	&	2.30$\pm$0.25	&1.97$\pm$0.09	&	16.19$\pm$0.24	&14.88$\pm$0.11	&	0.156	&0.063	&2.210$\pm$0.015 \\
WKK6233	&		12.73$\pm$0.07	&11.91$\pm$0.04	&	2.69$\pm$0.22	&2.14$\pm$0.07	&	16.66$\pm$0.19	&15.26$\pm$0.08	&	0.152	&0.062	&2.243$\pm$0.014 \\
WKK6235	&		11.91$\pm$0.06	&10.96$\pm$0.03	&	4.73$\pm$0.42	&4.39$\pm$0.21	&	17.07$\pm$0.20	&15.99$\pm$0.11	&	0.170	&0.069	&2.141$\pm$0.018 \\
WKK6242	&		11.82$\pm$0.07	&10.90$\pm$0.03	&	2.74$\pm$0.24	&2.16$\pm$0.15	&	15.86$\pm$0.20	&14.41$\pm$0.15	&	0.151	&0.061	&2.424$\pm$0.011 \\
WKK6250	&		11.56$\pm$0.05	&10.55$\pm$0.02	&	3.58$\pm$0.31	&3.41$\pm$0.13	&	16.15$\pm$0.20	&15.01$\pm$0.08	&	0.159	&0.065	&2.324$\pm$0.010 \\
WKK6252	&		13.00$\pm$0.06	&12.18$\pm$0.03	&	2.80$\pm$0.24	&2.44$\pm$0.16	&	17.04$\pm$0.19	&15.88$\pm$0.14	&	0.153	&0.063	&2.073$\pm$0.012 \\
WKK6269	&		8.97$\pm$0.06		&8.14$\pm$0.03	&	20.40$\pm$2.47&17.57$\pm$1.03	&	17.44$\pm$0.27	&16.22$\pm$0.13	&	0.148	&0.060	&2.579$\pm$0.011 \\
WKK6275	&		11.40$\pm$0.06	&10.22$\pm$0.03	&	7.74$\pm$0.78	&8.39$\pm$0.60	&	17.62$\pm$0.23	&16.68$\pm$0.16	&	0.149	&0.061	&2.165$\pm$0.014 \\
WKK6282	&		12.17$\pm$0.06	&11.22$\pm$0.03	&	3.05$\pm$0.33	&2.60$\pm$0.17	&	16.42$\pm$0.24	&15.14$\pm$0.15	&	0.147	&0.060	&2.274$\pm$0.012 \\
WKK6297	&		12.69$\pm$0.05	&11.78$\pm$0.02	&	4.38$\pm$0.25	&4.01$\pm$0.18	&	17.72$\pm$0.14	&16.71$\pm$0.10	&	0.149	&0.061	&1.844$\pm$0.026 \\
WKK6305	&		10.23$\pm$0.05	&9.24$\pm$0.11	&	6.77$\pm$0.79	&7.93$\pm$1.49	&	16.25$\pm$0.26	&15.45$\pm$0.42	&	0.198	&0.081	&2.327$\pm$0.009 \\
WKK6318	&		9.61$\pm$0.06		&8.73$\pm$0.03	&	15.84$\pm$1.76&13.75$\pm$0.79	&	17.49$\pm$0.25	&16.35$\pm$0.13	&	0.180	&0.073	&2.354$\pm$0.013 \\
WKK6342	&		12.08$\pm$0.05	&11.06$\pm$0.02	&	2.59$\pm$0.13	&3.05$\pm$0.15	&	15.98$\pm$0.12	&15.11$\pm$0.11	&	0.152	&0.062	&2.326$\pm$0.009 \\
WKK6360	&		10.92$\pm$0.06	&10.02$\pm$0.02	&	4.19$\pm$0.66	&3.60$\pm$0.15	&	15.91$\pm$0.35	&14.61$\pm$0.10	&	0.154	&0.063	&2.505$\pm$0.009 \\
WKK6383	&		11.94$\pm$0.06	&10.78$\pm$0.03	&	4.04$\pm$0.26	&4.59$\pm$0.22	&	16.77$\pm$0.15	&15.88$\pm$0.11	&	0.176	&0.072	&2.195$\pm$0.012 \\
WKK6402	&		11.56$\pm$0.05	&10.62$\pm$0.02	&	3.70$\pm$0.29	&3.23$\pm$0.15	&	16.20$\pm$0.18	&14.98$\pm$0.10	&	0.190	&0.077	&2.238$\pm$0.011 \\
WKK6429	&		12.30$\pm$0.04	&11.63$\pm$0.04	&	5.04$\pm$0.37	&5.56$\pm$0.17	&	17.73$\pm$0.16	&16.91$\pm$0.09	&	0.159	&0.065	&2.084$\pm$0.012 \\
WKK6431	&		11.74$\pm$0.05	&10.78$\pm$0.03	&	3.56$\pm$0.24	&3.18$\pm$0.19	&	16.28$\pm$0.16	&15.17$\pm$0.13	&	0.184	&0.075	&2.286$\pm$0.010 \\
WKK6459	&		11.37$\pm$0.06	&10.45$\pm$0.04	&	6.18$\pm$0.85	&6.25$\pm$0.21	&	17.09$\pm$0.31	&16.13$\pm$0.08	&	0.200	&0.081	&2.322$\pm$0.010 \\
WKK6477	&		12.67$\pm$0.05	&11.73$\pm$0.04	&	3.02$\pm$0.32	&2.68$\pm$0.13	&	16.90$\pm$0.23	&15.76$\pm$0.11	&	0.179	&0.073	&2.106$\pm$0.019 \\
%
\hline
\end{tabular}
\end{table*}

\subsection{
Fundamental Plane -- Fit results}\label{JKs_fittings}

\begin{figure}[htb!]
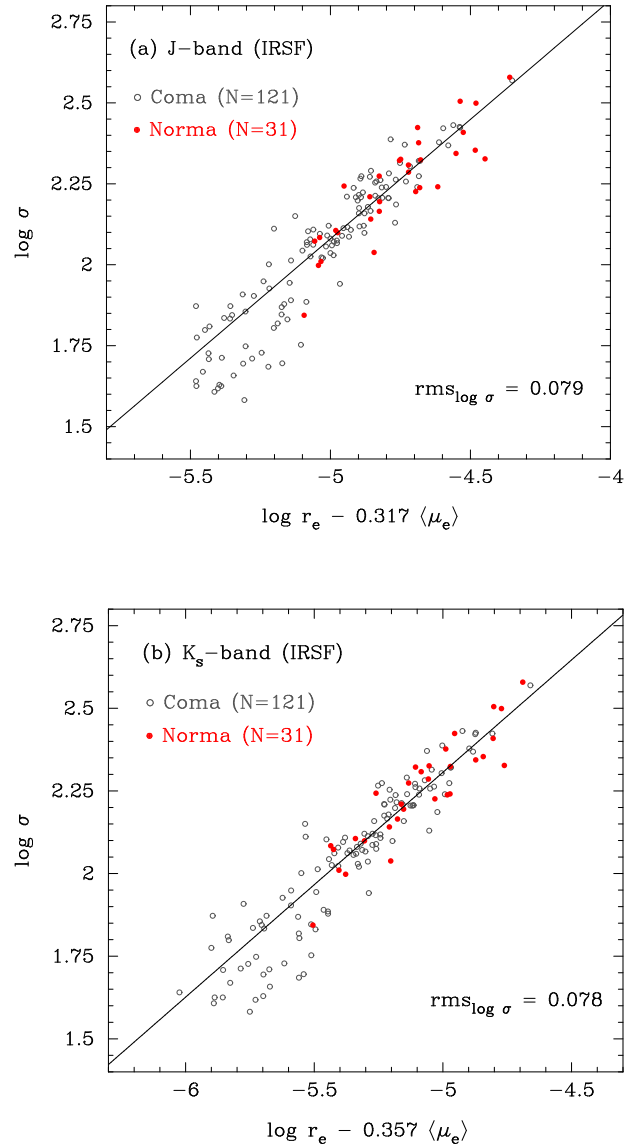

 \centering
 \begin{tabular}{c}
\includegraphics[width=0.45\textwidth]{FP_IRSF-Jband.eps} \\
~\\~\\
\includegraphics[width=0.45\textwidth]{FP_IRSF-Ksband.eps}
\end{tabular}
\caption{\label{JKs:fig_FP}The $J$ and $K_s$-band FP based on 31 ETGs from the IRSF data (for the Norma cluster) and 121 ETGs from 2MASS (for the Coma calibration sample). The top (a) and bottom (b) panels 
represent the $J$- and $K_s$-band FP projections, respectively. The red filled circles represent the Norma cluster ETGs after shifting them to the Coma cluster's distance. The black open circles represent the Coma cluster galaxies.}
\end{figure}

The FP was fitted using the inverse least-squares fit. The top and bottom 
panels of \fref{JKs:fig_FP} show the FP projection for the $J$- and $K_s$-band, respectively. The black open circles represent the Coma cluster galaxies while the red filled circles are the Norma cluster galaxies shifted to the Coma distance. This was done assuming the Coma cluster has a zero peculiar velocity. %

\begin{figure}[htb!]
 \centering
 \begin{tabular}{c}
\includegraphics[width=0.45\textwidth]{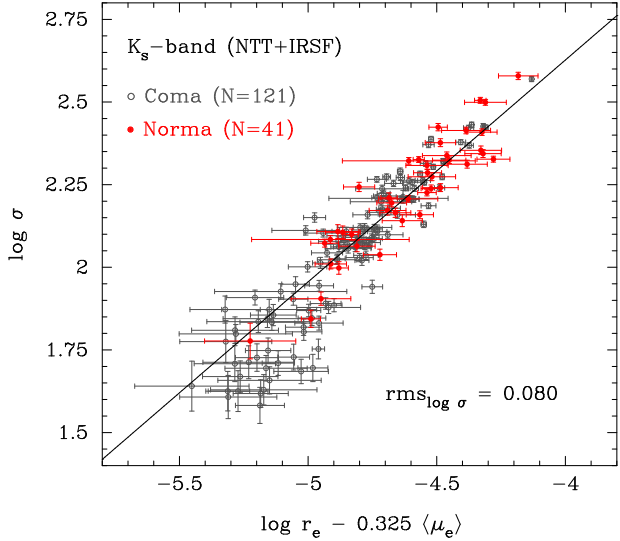}
\end{tabular}
\caption[Combined $K_s$-band FP]{\label{Ks:fig_FP}Same as \fref{JKs:fig_FP} but for the combined $K_s$-band FP projection. 
}
\end{figure}

The combined $K_s$-band FP analysis was performed using the combined IRSF and NTT data. In combining the two data sets, the latter were given preference due to the higher resolution imaging. The FP projection obtained from the combined $K_s$-band analysis is shown in \fref{Ks:fig_FP}. The black open circles represent the Coma cluster galaxies while the red filled circles represent the Norma cluster galaxies after shifting them to the Coma cluster's distance. 
A summary of results for the IRSF $J$ and $K_s$-band as well as the combined $K_s$-band FP fit parameters is presented in \tref{JKs_FP_distance}. The derived values for the FP parameters $a$, and $b$ are comparable with other previous FP studies in the near-infrared \citep[see e.g.,][]{LaBarbera_10, Magoulas_12}. %

\begin{table} \tiny 
\centering
\caption[$J$- and $K_s$-band FP fit parameters]{\label{JKs_FP_distance}The $J$- and $K_s$-band FP fit parameters ($a$ and $b$) as well as the zero-point offset. 
The first column indicates the source of the data used while 
column~5 is the FP zero-point offset. Columns 6 and 7 are the scatter and the corresponding percentage uncertainty in the distance to the individual Norma cluster galaxies, respectively. Column 8 indicates the number of Norma cluster galaxies used in our analysis; this is 121 for the Coma cluster.}
\begin{tabular}{l@{\hskip 0.075in}c@{\hskip 0.075in}c@{\hskip 0.075in}c@{\hskip 0.075in}c@{\hskip 0.075in}c@{\hskip 0.075in}c@{\hskip 0.075in}l} \hline\hline
Sample  &  Filter & $a$ &                           $b$ &                         $\Delta c$ &        rms      &  \% err  &  N  \\ 
  &   &  &                           &                         &             &                &    \\ 
(1)    &   (2)    &  (3)    &      (4)                    &          (5)               &      (6)    &      (7)         &          (8)   \\ \hline
%
%
NTT   &    $K_s$ & $1.465\pm0.059$ &  $0.326\pm0.020$ &  $0.154\pm0.014$ &  0.083     &  28  &  31$*$  \\
IRSF  &    $J$   & $1.356\pm0.058$ &  $0.317\pm0.018$ &  $0.149\pm0.016$ &  0.079     &  25  &  31  \\
IRSF  &    $K_s$ & $1.470\pm0.053$ &  $0.357\pm0.019$ &  $0.145\pm0.014$ &  0.078     &  26  &  31  \\
NTT+IRSF & $K_s$ & $1.490\pm0.053$ &  $0.325\pm0.017$ &  $0.163\pm0.013$ &  0.080     &  27  &  41  \\
\hline
\end{tabular}
~\\$*$ --- All data in this row were taken from \citet{Mutabazi_14}.
\end{table}

\section{Norma cluster: distance and peculiar velocity} \label{dist_vpec_fp} 

\tref{JKs_FP_distance} shows the different zero-point offsets measured using the $J$ and $K_s$-band FP analysis. These are in excellent agreement with each other. The best measured value (least scatter/error) is from the combined $K_s$-band FP analysis using 41 ETGs within the Norma cluster. We adopted this value to measure the distance and peculiar velocity for the Norma cluster. This implies that 
\begin{equation}
\log {D_A \left(\rm{Coma}\right)} - \log{D_A\left(\rm{Norma}\right)} =
0.163\,\pm\,0.013,
\label{finedist_here}
\end{equation}
where $D_A (\rm{Coma})$ and $D_A (\rm{Norma})$ are the angular diameter distances for the Coma and Norma clusters, respectively. The angular diameter distances were measured following \citet{Wright_06}. The distance (Hubble redshift, $z_{_{\rm H}}$) and the peculiar velocity for the Norma cluster were derived using the same method described in \citet{Mutabazi_14}. That is, the peculiar velocity for the Norma cluster was computed using 
\begin{equation}
 v_{\text{pec}} = c\,z_{\text{pec}}, 
\end{equation}
where $c$ is the speed of light and 
\begin{equation}
 z_{\text{pec}} = \frac{1+z_{_{\text{CMB}}}}{1+z_{_{\text{H}}}} - 1. 
\end{equation}
The resulting Norma cluster's peculiar velocity after applying the homogeneous Malmquist bias correction \citep[see][for details]{Hudson_97} presented in Table~\ref{JKs_FP_distance_vpec} is $44\pm151$\;km\;s$^{-1}$ (distance of 4915\,$\pm$\,121\;km\;s$^{-1}$ or $d$\,$=$\,69.7\,$\pm$\,1.7~Mpc). 
\begin{table}  \tiny 
\centering
\caption[Distance and peculiar velocity from the $J$- and $K_s$-band FP analysis]{\label{JKs_FP_distance_vpec}The $J$- and $K_s$-band FP offsets (column~3) and the corresponding measured peculiar velocity (column~5). The homogeneous Malmquist bias correction shown in column 6 has not been applied (this should be subtracted from the peculiar velocity presented in column~5). %
}
\begin{tabular}{l@{\hskip 0.1in}c@{\hskip 0.1in}c@{\hskip 0.1in}c@{\hskip 0.1in}r@{\hskip 0.1in}c@{\hskip 0.1in}l} \hline\hline
Sample  &  Filter &   $\Delta c$ &                  $z_{_{\rm H}}$ &            \multicolumn{1}{c@{\hskip 0.1in}}{v$_{\rm{pec}}$} &    Malm. bias &  N  \\ 
  &                  &                               [dex]      &        &          [km\;s$^{-1}$]     &               [km\;s$^{-1}$]   &    \\ 
 (1)   &  (2)        &          (3)               &        (4)           &    \multicolumn{1}{c@{\hskip 0.1in}}{(5)}      &   (6)           &  (7)    \\ \hline
%
%
NTT       &  $K_s$  & $0.154\pm0.014$ & $0.01667\pm0.00055$  & $-43\pm170$   &  29      &  31$*$  \\
IRSF      &  $J$    & $0.149\pm0.016$ & $0.01684\pm0.00062$  & $-95\pm187$   &  33      &  31  \\
IRSF      &  $K_s$  & $0.145\pm0.014$ & $0.01704\pm0.00058$  & $-152\pm173$  &  30      &  31  \\
NTT+IRSF  &  $K_s$  & $0.163\pm0.013$ & $0.01632\pm0.00050$  & $+58\pm151$   &  14      &  41  \\
\hline
\end{tabular}
~\\$*$ --- All data in this row were taken from \citet{Mutabazi_14}.
\end{table}
%
%
%

\section{Discussion and conclusions} \label{sec_discuss}

We have measured a small peculiar velocity of the Norma cluster of $+44$\,$\pm$\,$151$\;km\;s$^{-1}$ which within errors, is consistent with zero and in excellent agreement with our previous measurement of $-72\pm170$\;km\;s$^{-1}$ \citep{Mutabazi_14}. Our analysis and results have demonstrated, that, despite the challenges of the large Galactic extinction and severe stellar contamination, distances using the FP and TF relations can be derived reliably for galaxies that lie relatively close to the Galactic plane. 
We have carefully considered the possible sources of measurement and systematic errors. Results from simulations of ETGs showed that the ef\mbox{}fect of star-subtraction is very small, ranging from 
$-0\!\fmg 011$ (NTT $K_s$-band data) to 
$-0\!\fmg 014$ (IRSF $K_s$-band data), which offsets the measured peculiar velocity by $\sim$\,41\;km\;s$^{-1}$ to $\sim$\,52\;km\;s$^{-1}$, respectively. The systematic effect arising from possible gradients in the sky background 
is very small, i.e., $\sim$\,$-0\!\fmg 008$, which corresponds to a change in Norma's measured peculiar velocity of $\sim$\,30\;km\;s$^{-1}$. %

Using the NIR $J$ and $K_s$-band images obtained using the IRSF telescope has also helped re-calibrate the \citet{Schlegel_98} extinction values as described in \Sref{extinc_recalib}. For the Norma cluster sample, the measured extinction correction factor is 0.81 for the \citeauthor{Schlegel_98} maps, which is in good agreement with the \citet{Schlafly_11} extinction values. The combined $K_s$-band FP sample resulted in the largest cluster sample ($N=41$) used in the peculiar velocity studies at relatively low Galactic latitudes to date, and hence the most precise distance and peculiar velocity of the Norma cluster measured to date. 

\begin{figure}[h]
 \centering 
\includegraphics[width=0.43\textwidth]{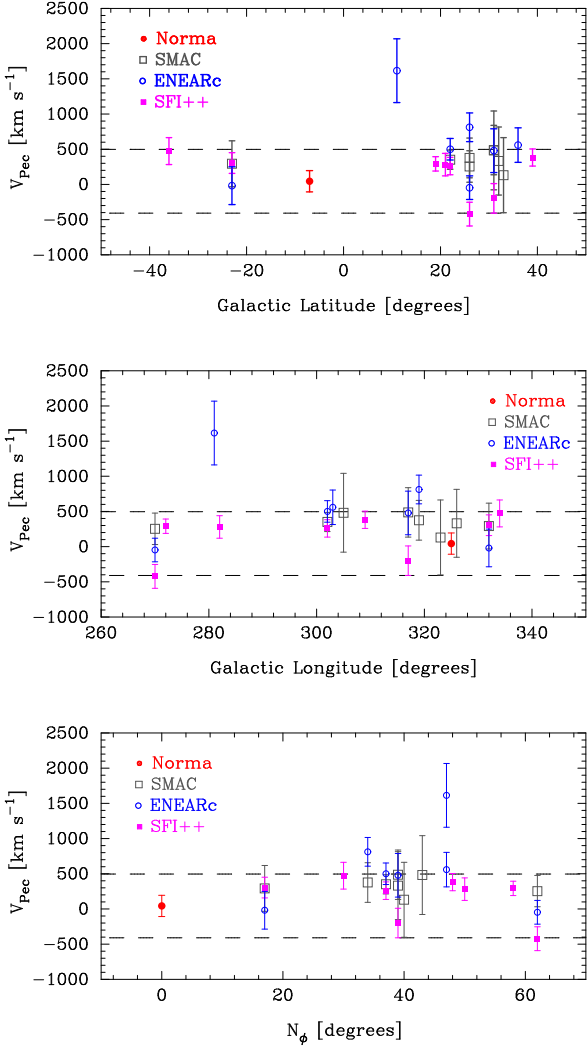}  
\caption{\label{GA_vpec_plot}
Peculiar velocities for Clusters/Groups in the GA region. The Norma cluster is represented by the red filled circle. The dashed horizontal lines represent $\pm3\sigma$ where $\sigma$ is the error on the Norma cluster's peculiar velocity measured using the combined $K_s$-band FP analysis. $N_{\phi}$ is the angle on the sky between the cluster and Norma. %
}
\end{figure}

\fref{GA_vpec_plot} shows the peculiar velocities for Clusters/Groups in the GA region. The peculiar velocity measurements were taken from the literature. 
Shown are the velocity measurements from ENEARc \citep{Bernardi_02}, SMAC \citep{Hudson_04}, SFI++ \citep{Springob_07}, and our Norma cluster measurement (red filled circle). The dashed horizontal lines represent $\pm 3\sigma$ where $\sigma$\,$=$\,151\;km\;s$^{-1}$ is the error on the Norma cluster's peculiar velocity derived using the combined $K_s$-band FP analysis. $N_{\phi}$ is the angular separation between the cluster and Norma. %
The improved sample of 41 ETGs taken from within $\frac{2}{3}$ of the Norma cluster's Abell radius makes this the most precise measurement for the distance and peculiar velocity of the Norma cluster to date. %

In the study by \citet{Carrick_15} the resisual bulk flow due to structures not covered by the 2M++ redshift sample has a velocity of $159\pm 23$~km\;s$^{-1}$ in the direction $\ell = 303^\circ \pm 11^\circ$, $b = 6^\circ \pm 11^\circ$. This is $\sim 25^\circ$ from Norma, and therefore 
corresponds to $\sim 144\pm 21$~km\;s$^{-1}$. This is the external bulk flow contribution on the Norma peculiar velocity arising from structures beyond the volume covered by the 2M++ redshift catalogue \citep{Lavaux_2011}. Similar residual bulk flows have been reported with a bulk flow contribution on the Norma cluster of e.g., $136\pm39$~km\;s$^{-1}$ \citep{Turnbull_12}, $154\pm10$~km\;s$^{-1}$ \citep{Boruah_19}, $147\pm12$~km\;s$^{-1}$ \citep{Said_20}, and $161\pm11$~km\;s$^{-1}$ \citep{Said_20}. These are all significantly different from zero and are, within the error margin ($1\sigma$), consistent with our Norma's peculiar velocity of $44$\,$\pm$\,$151$\;km\;s$^{-1}$. Dedicated surveys such as Taipan \citep{Cunha_17, Taylor_20} and WALLABY \citep{Duffy_12, Koribalski_20} will help improve the accuracy of distance measurements and hence enable us infer more accurate peculiar velocities. 

\acknowledgments
TM acknowledges financial support from the Square Kilometre Array South Africa (SKA~SA). Travel costs to Sutherland site of the South African Astronomical Observatory were funded by the multi-wavelength funding (Department of Astronomy, University of Cape Town). 
Funding for SDSS-III has been provided by the Alfred P. Sloan Foundation, the Participating Institutions, the National Science Foundation, and the U.S. Department of Energy Office of Science. The SDSS-III web site is \url{http://www.sdss3.org/}. SDSS-III is managed by the Astrophysical Research Consortium for the Participating Institutions of the SDSS-III Collaboration including the University of Arizona, the Brazilian Participation Group, Brookhaven National Laboratory, Carnegie Mellon University, University of Florida, the French Participation Group, the German Participation Group, Harvard University, the Instituto de Astrofisica de Canarias, the Michigan State/Notre Dame/JINA Participation Group, Johns Hopkins University, Lawrence Berkeley National Laboratory, Max Planck Institute for Astrophysics, Max Planck Institute for Extraterrestrial Physics, New Mexico State University, New York University, Ohio State University, Pennsylvania State University, University of Portsmouth, Princeton University, the Spanish Participation Group, University of Tokyo, University of Utah, Vanderbilt University, University of Virginia, University of Washington, and Yale University. This publication has made use of the NASA/IPAC ExtraGalactic Data base (NED), and also data products from the 2MASS, a joint project of the University of Massachusetts and the Infrared Processing and Analysis Center California Institute of Technology, funded by the National Aeronautics and Space Administration and the National Science Foundation.
%


\software
\galfit\ \citep{Peng_10}, \matplotlib\ \citep{Hunter_07}, \scipy\ \citep{Virtanen_17}, \numpy\ \citep{Oliphant_06, 
van_11}, \pgplot\ \citep{Pearson_11}, \ellipse\ task under \iraf's \stsdas\ package \citep{Jedrzejewski_87}.

\facilities{1.4m Infrared Survey Facility (IRSF) at South Africa Astronomical Observatory Sutherland site,
European Southern Observatory (ESO) 3.6m New Techology Telescope (NTT) at La Silla Observatory,
Australian National University (ANU) 3.9m Anglo-Australian Telescope (AAT) at Siding Spring Observatory (SSO).}

\bibliographystyle{aasjournal} 
\bibliography{mn}

\appendix


\section{Coma cluster ETGs -- Calibration Sample} \label{sec:appendix}
The Coma cluster galaxies used in our calibration sample are those which (1) are classified as E/S0 or S0 in \citet{Dressler_80} (2) are Coma cluster members based on their redshifts, (3) have central velocity dispersions available in SDSS DR8 \citep{Aihara_11}, and (4) we were able to successfully measure their total extrapolated magnitudes and effective radii using the 2MASS Extended Source \citep{Skrutskie_06} Atlas images. The photometry measurements (total extrapolated magnitudes, effective radii, and the computed mean effective surface brightness) in the $J$ and $K_s$-bands) obtained from surface brightness profile fitting of the Coma cluster ETGs are summarised in Table \ref{Coma_JKs_phot}.
\begin{table*} 
\begin{center}
\caption[Coma cluster photometry results]{\label{Coma_JKs_phot} Coma cluster $J$ and $K_s$ photometry results from f\mbox{}itting and extrapolating the galaxy surface brightness prof\mbox{}iles using the 2MASS Atlas images. No corrections have been applied to the total magnitudes presented here.}
\begin{tabular}{c@{\hskip 0.15in} r@{\hskip 0.15in} r@{\hskip 0.15in} r@{\hskip 0.15in} r@{\hskip 0.15in} c@{\hskip 0.15in} c@{\hskip 0.15in} c}  \\ \hline\hline
\multicolumn{1}{c}{Identification}  & \multicolumn{2}{c@{\hskip 0.15in}}{Total magnitude} &  \multicolumn{2}{c@{\hskip 0.15in}}{Effective radius} &  \multicolumn{2}{c@{\hskip 0.15in}}{Mean eff. surf. brightness} &  \multicolumn{1}{c}{$\log \sigma$} \\ 
 & \multicolumn{1}{c@{\hskip 0.15in}}{$J$}  & \multicolumn{1}{c@{\hskip 0.15in}}{$K_s$}  & \multicolumn{1}{c@{\hskip 0.15in}}{$J$} & \multicolumn{1}{c@{\hskip 0.15in}}{$K_s$} & \multicolumn{1}{c@{\hskip 0.15in}}{$J$} & \multicolumn{1}{c@{\hskip 0.15in}}{$K_s$}  &         \\ 
\multicolumn{1}{c@{\hskip 0.15in}}{(1)} & \multicolumn{1}{c@{\hskip 0.15in}}{(2)}  & \multicolumn{1}{c@{\hskip 0.15in}}{(3)}  & \multicolumn{1}{c@{\hskip 0.15in}}{(4)} & \multicolumn{1}{c@{\hskip 0.15in}}{(5)} & \multicolumn{1}{c@{\hskip 0.15in}}{(6)} & \multicolumn{1}{c@{\hskip 0.15in}}{(7)}   & \multicolumn{1}{c}{(8)}  \\ \hline
2MASXJ13023273+2717443	&		14.15$\pm$0.08	&13.48$\pm$0.15	&	3.98$\pm$0.21	&3.23$\pm$0.59	&	19.05$\pm$0.14	&17.83$\pm$0.42	&	1.709$\pm$0.055 \\
2MASXJ13020552+2717499	&		13.93$\pm$0.06	&13.51$\pm$0.14	&	4.19$\pm$0.08	&2.21$\pm$0.52	&	18.95$\pm$0.07	&17.04$\pm$0.53	&	1.836$\pm$0.033 \\
2MASXJ13000623+2718022	&		13.63$\pm$0.04	&12.82$\pm$0.07	&	4.59$\pm$0.18	&4.46$\pm$0.46	&	18.83$\pm$0.09	&17.86$\pm$0.23	&	1.748$\pm$0.038 \\
2MASXJ12593730+2720097	&		14.74$\pm$0.11	&14.27$\pm$0.24	&	2.25$\pm$0.17	&2.29$\pm$0.58	&	18.41$\pm$0.20	&17.89$\pm$0.60	&	1.640$\pm$0.076 \\
2MASXJ13010615+2723522	&		14.31$\pm$0.09	&13.83$\pm$0.17	&	3.26$\pm$0.44	&2.15$\pm$0.41	&	18.76$\pm$0.31	&17.28$\pm$0.45	&	1.809$\pm$0.041 \\
2MASXJ12564777+2725158	&		14.25$\pm$0.10	&13.48$\pm$0.12	&	3.11$\pm$0.33	&2.24$\pm$0.27	&	18.61$\pm$0.25	&17.03$\pm$0.29	&	1.618$\pm$0.042 \\
2MASXJ12583209+2727227	&		13.38$\pm$0.05	&12.57$\pm$0.05	&	2.00$\pm$0.10	&1.87$\pm$0.20	&	16.78$\pm$0.11	&15.74$\pm$0.23	&	2.056$\pm$0.014 \\
2MASXJ12573614+2729058	&		13.15$\pm$0.03	&12.15$\pm$0.04	&	1.97$\pm$0.04	&2.05$\pm$0.04	&	16.53$\pm$0.05	&15.52$\pm$0.06	&	2.210$\pm$0.009 \\
2MASXJ12570940+2727587	&		12.24$\pm$0.02	&11.16$\pm$0.02	&	2.55$\pm$0.32	&3.02$\pm$0.02	&	16.17$\pm$0.27	&15.37$\pm$0.02	&	2.304$\pm$0.008 \\
2MASXJ13002689+2730556	&		13.49$\pm$0.05	&12.52$\pm$0.05	&	1.86$\pm$0.09	&2.40$\pm$0.21	&	16.74$\pm$0.11	&16.21$\pm$0.20	&	2.023$\pm$0.014 \\
2MASXJ12580974+2732585	&		14.25$\pm$0.08	&13.28$\pm$0.10	&	3.48$\pm$0.30	&2.95$\pm$0.27	&	18.87$\pm$0.20	&17.45$\pm$0.22	&	1.727$\pm$0.042 \\
2MASXJ12573584+2729358	&		11.97$\pm$0.02	&10.90$\pm$0.02	&	3.07$\pm$0.05	&3.57$\pm$0.02	&	16.31$\pm$0.04	&15.47$\pm$0.02	&	2.321$\pm$0.009 \\
2MASXJ12572435+2729517	&		9.85$\pm$0.00	&9.07$\pm$0.00	&	27.31$\pm$0.05	&21.19$\pm$0.20	&	18.93$\pm$0.00	&17.51$\pm$0.02	&	2.431$\pm$0.008 \\
2MASXJ12570431+2731328	&		14.24$\pm$0.08	&13.63$\pm$0.15	&	4.32$\pm$0.48	&3.17$\pm$0.49	&	19.31$\pm$0.25	&17.92$\pm$0.37	&	1.872$\pm$0.041 \\
2MASXJ12563418+2732200	&		12.71$\pm$0.03	&11.61$\pm$0.03	&	2.80$\pm$0.08	&3.59$\pm$0.07	&	16.85$\pm$0.07	&16.20$\pm$0.05	&	2.208$\pm$0.009 \\
2MASXJ13014841+2736147	&		13.51$\pm$0.04	&12.82$\pm$0.07	&	3.21$\pm$0.16	&2.41$\pm$0.14	&	17.94$\pm$0.12	&16.51$\pm$0.15	&	1.846$\pm$0.022 \\
2MASXJ13011224+2736162	&		13.76$\pm$0.06	&12.53$\pm$0.05	&	3.09$\pm$0.15	&4.37$\pm$0.31	&	18.11$\pm$0.12	&17.53$\pm$0.16	&	1.729$\pm$0.039 \\
2MASXJ13001914+2733135	&		12.88$\pm$0.03	&11.79$\pm$0.03	&	3.28$\pm$0.08	&3.90$\pm$0.24	&	17.37$\pm$0.06	&16.59$\pm$0.14	&	2.030$\pm$0.013 \\
2MASXJ12585812+2735409	&		13.16$\pm$0.06	&12.06$\pm$0.04	&	1.37$\pm$0.22	&2.12$\pm$0.10	&	15.76$\pm$0.35	&15.53$\pm$0.11	&	2.166$\pm$0.011 \\
2MASXJ12573284+2736368	&		11.65$\pm$0.01	&10.62$\pm$0.01	&	3.47$\pm$0.06	&3.76$\pm$0.09	&	16.27$\pm$0.04	&15.33$\pm$0.05	&	2.379$\pm$0.008 \\
2MASXJ13020106+2739109	&		13.41$\pm$0.03	&12.70$\pm$0.07	&	4.01$\pm$0.19	&3.02$\pm$0.24	&	18.32$\pm$0.11	&16.91$\pm$0.19	&	1.805$\pm$0.026 \\
2MASXJ13015375+2737277	&		10.90$\pm$0.01	&10.05$\pm$0.01	&	6.76$\pm$0.20	&5.49$\pm$0.07	&	16.94$\pm$0.06	&15.54$\pm$0.03	&	2.423$\pm$0.008 \\
2MASXJ12591030+2737119	&		13.28$\pm$0.04	&12.34$\pm$0.05	&	2.76$\pm$0.18	&2.75$\pm$0.25	&	17.40$\pm$0.15	&16.38$\pm$0.20	&	2.078$\pm$0.014 \\
2MASXJ12571682+2737068	&		13.51$\pm$0.06	&12.48$\pm$0.05	&	1.12$\pm$0.28	&1.18$\pm$0.33	&	15.65$\pm$0.55	&14.64$\pm$0.61	&	2.208$\pm$0.009 \\
2MASXJ12594713+2742372	&		12.05$\pm$0.05	&11.14$\pm$0.02	&	4.04$\pm$0.27	&3.66$\pm$0.04	&	16.97$\pm$0.15	&15.74$\pm$0.03	&	2.130$\pm$0.010 \\
2MASXJ12584742+2740288	&		11.97$\pm$0.01	&11.07$\pm$0.02	&	7.20$\pm$0.15	&6.77$\pm$0.11	&	18.14$\pm$0.05	&17.00$\pm$0.04	&	2.273$\pm$0.009 \\
2MASXJ12583157+2740247	&		13.94$\pm$0.06	&12.90$\pm$0.07	&	1.27$\pm$0.15	&1.80$\pm$0.15	&	16.37$\pm$0.26	&15.99$\pm$0.20	&	2.104$\pm$0.012 \\
2MASXJ12563420+2741150	&		14.28$\pm$0.07	&13.82$\pm$0.21	&	3.54$\pm$0.37	&2.09$\pm$0.53	&	18.93$\pm$0.24	&17.24$\pm$0.59	&	1.798$\pm$0.029 \\
2MASXJ13000626+2746332	&		12.97$\pm$0.03	&11.98$\pm$0.04	&	4.29$\pm$0.25	&4.21$\pm$0.15	&	18.04$\pm$0.13	&16.94$\pm$0.09	&	2.061$\pm$0.013 \\
2MASXJ12592491+2744198	&		13.15$\pm$0.04	&12.27$\pm$0.04	&	1.65$\pm$0.76	&1.96$\pm$0.13	&	16.15$\pm$1.00	&15.58$\pm$0.15	&	2.159$\pm$0.012 \\
2MASXJ12591348+2746289	&		12.65$\pm$0.02	&11.90$\pm$0.03	&	4.39$\pm$0.20	&3.41$\pm$0.11	&	17.77$\pm$0.10	&16.38$\pm$0.08	&	2.071$\pm$0.014 \\
2MASXJ12590821+2747029	&		12.12$\pm$0.03	&11.13$\pm$0.02	&	3.38$\pm$0.09	&3.45$\pm$0.08	&	16.67$\pm$0.07	&15.63$\pm$0.06	&	2.314$\pm$0.009 \\
2MASXJ12590745+2746039	&		12.90$\pm$0.05	&11.98$\pm$0.04	&	2.13$\pm$0.10	&2.25$\pm$0.06	&	16.45$\pm$0.11	&15.57$\pm$0.07	&	2.223$\pm$0.010 \\
2MASXJ12585766+2747079	&		13.69$\pm$0.05	&13.12$\pm$0.09	&	4.24$\pm$0.40	&3.41$\pm$0.37	&	18.73$\pm$0.21	&17.60$\pm$0.25	&	1.582$\pm$0.055 \\
2MASXJ12585208+2747059	&		12.43$\pm$0.02	&11.65$\pm$0.03	&	4.01$\pm$0.24	&3.34$\pm$0.09	&	17.37$\pm$0.13	&16.11$\pm$0.06	&	2.175$\pm$0.011 \\
2MASXJ12581922+2745437	&		14.28$\pm$0.08	&13.48$\pm$0.13	&	2.41$\pm$0.12	&1.86$\pm$0.40	&	18.12$\pm$0.13	&16.68$\pm$0.49	&	1.872$\pm$0.031 \\
2MASXJ12574616+2745254	&		12.75$\pm$0.02	&12.15$\pm$0.04	&	7.93$\pm$0.20	&5.02$\pm$0.34	&	19.17$\pm$0.06	&17.49$\pm$0.15	&	1.696$\pm$0.042 \\
2MASXJ13011761+2748321	&		11.91$\pm$0.02	&10.88$\pm$0.01	&	3.38$\pm$0.03	&3.69$\pm$0.03	&	16.46$\pm$0.02	&15.53$\pm$0.02	&	2.274$\pm$0.009 \\
2MASXJ13003334+2749266	&		14.46$\pm$0.11	&13.53$\pm$0.12	&	3.25$\pm$0.41	&3.44$\pm$0.31	&	18.91$\pm$0.30	&18.00$\pm$0.23	&	1.625$\pm$0.060 \\
2MASXJ13000551+2748272	&		12.55$\pm$0.02	&11.89$\pm$0.03	&	4.75$\pm$0.16	&3.30$\pm$0.10	&	17.84$\pm$0.08	&16.31$\pm$0.07	&	2.067$\pm$0.013 \\
2MASXJ12595489+2747453	&		14.40$\pm$0.11	&13.74$\pm$0.13	&	2.07$\pm$0.35	&1.47$\pm$0.35	&	17.87$\pm$0.39	&16.36$\pm$0.53	&	1.658$\pm$0.041 \\
2MASXJ12593697+2749327	&		14.50$\pm$0.12	&13.62$\pm$0.15	&	1.08$\pm$0.10	&1.36$\pm$0.41	&	16.57$\pm$0.23	&16.13$\pm$0.68	&	1.927$\pm$0.023 \\
2MASXJ12592936+2751008	&		12.35$\pm$0.03	&11.41$\pm$0.02	&	2.87$\pm$0.26	&2.29$\pm$0.02	&	16.55$\pm$0.20	&15.02$\pm$0.03	&	2.387$\pm$0.009 \\
2MASXJ12580349+2748535	&		12.85$\pm$0.04	&12.03$\pm$0.04	&	2.44$\pm$0.10	&2.03$\pm$0.14	&	16.69$\pm$0.10	&15.37$\pm$0.15	&	2.216$\pm$0.009 \\
2MASXJ12574728+2749594	&		12.89$\pm$0.03	&12.11$\pm$0.04	&	2.25$\pm$0.15	&2.16$\pm$0.05	&	16.57$\pm$0.14	&15.63$\pm$0.06	&	2.118$\pm$0.012 \\
2MASXJ12571778+2748388	&		13.89$\pm$0.07	&12.86$\pm$0.08	&	2.98$\pm$0.25	&3.62$\pm$0.25	&	18.16$\pm$0.19	&17.46$\pm$0.17	&	1.710$\pm$0.038 \\
  \hline
\end{tabular}
\end{center}
\end{table*}      

\begin{table*} 
\begin{center}
Table \ref{Coma_JKs_phot} continued \ldots
\begin{tabular}{c@{\hskip 0.15in} r@{\hskip 0.15in} r@{\hskip 0.15in} r@{\hskip 0.15in} r@{\hskip 0.15in} c@{\hskip 0.15in} c@{\hskip 0.15in} c}  \\ \hline\hline
\multicolumn{1}{c}{Identification}  & \multicolumn{2}{c@{\hskip 0.15in}}{Total magnitude} &  \multicolumn{2}{c@{\hskip 0.15in}}{Effective radius} &  \multicolumn{2}{c@{\hskip 0.15in}}{Mean eff. surf. brightness} &  \multicolumn{1}{c}{$\log \sigma$} \\ 
 & \multicolumn{1}{c@{\hskip 0.15in}}{$J$}  & \multicolumn{1}{c@{\hskip 0.15in}}{$K_s$}  & \multicolumn{1}{c@{\hskip 0.15in}}{$J$} & \multicolumn{1}{c@{\hskip 0.15in}}{$K_s$} & \multicolumn{1}{c@{\hskip 0.15in}}{$J$} & \multicolumn{1}{c@{\hskip 0.15in}}{$K_s$}  &         \\ 
\multicolumn{1}{c@{\hskip 0.15in}}{(1)} & \multicolumn{1}{c@{\hskip 0.15in}}{(2)}  & \multicolumn{1}{c@{\hskip 0.15in}}{(3)}  & \multicolumn{1}{c@{\hskip 0.15in}}{(4)} & \multicolumn{1}{c@{\hskip 0.15in}}{(5)} & \multicolumn{1}{c@{\hskip 0.15in}}{(6)} & \multicolumn{1}{c@{\hskip 0.15in}}{(7)}   & \multicolumn{1}{c}{(8)}  \\ \hline
2MASXJ13025272+2751593	&		12.25$\pm$0.02	&11.29$\pm$0.02	&	3.10$\pm$0.35	&2.82$\pm$0.10	&	16.59$\pm$0.25	&15.33$\pm$0.08	&	2.186$\pm$0.009 \\
2MASXJ13015023+2753367	&		12.52$\pm$0.02	&11.67$\pm$0.03	&	3.09$\pm$0.08	&2.64$\pm$0.02	&	16.88$\pm$0.06	&15.58$\pm$0.03	&	2.290$\pm$0.010 \\
2MASXJ12594610+2751257	&		13.22$\pm$0.04	&12.08$\pm$0.04	&	1.97$\pm$0.04	&2.48$\pm$0.08	&	16.58$\pm$0.06	&15.84$\pm$0.08	&	2.090$\pm$0.011 \\
2MASXJ12593789+2754267	&		12.64$\pm$0.03	&11.49$\pm$0.03	&	2.36$\pm$0.02	&2.93$\pm$0.11	&	16.40$\pm$0.04	&15.61$\pm$0.09	&	2.261$\pm$0.009 \\
2MASXJ12592016+2753098	&		13.41$\pm$0.04	&12.48$\pm$0.05	&	3.28$\pm$0.20	&3.12$\pm$0.35	&	17.90$\pm$0.14	&16.78$\pm$0.25	&	1.831$\pm$0.021 \\
2MASXJ12590459+2754389	&		13.33$\pm$0.04	&12.73$\pm$0.08	&	2.94$\pm$0.14	&2.06$\pm$0.13	&	17.58$\pm$0.11	&16.13$\pm$0.16	&	2.044$\pm$0.012 \\
2MASXJ12590791+2751179	&		12.45$\pm$0.03	&11.46$\pm$0.03	&	3.11$\pm$0.20	&2.78$\pm$0.11	&	16.82$\pm$0.15	&15.51$\pm$0.09	&	2.258$\pm$0.009 \\
2MASXJ12575059+2752454	&		13.49$\pm$0.03	&12.50$\pm$0.05	&	2.82$\pm$0.65	&3.03$\pm$0.10	&	17.65$\pm$0.50	&16.72$\pm$0.08	&	2.013$\pm$0.017 \\
2MASXJ12572169+2752498	&		13.89$\pm$0.06	&12.92$\pm$0.09	&	2.38$\pm$0.06	&2.44$\pm$0.30	&	17.67$\pm$0.08	&16.66$\pm$0.28	&	1.685$\pm$0.037 \\
2MASXJ13004285+2757476	&		13.25$\pm$0.07	&12.11$\pm$0.04	&	2.21$\pm$0.25	&3.05$\pm$0.22	&	16.85$\pm$0.25	&16.31$\pm$0.17	&	2.060$\pm$0.012 \\
2MASXJ13004737+2755196	&		13.02$\pm$0.03	&12.54$\pm$0.06	&	5.33$\pm$0.25	&3.09$\pm$0.20	&	18.54$\pm$0.11	&16.76$\pm$0.15	&	1.944$\pm$0.016 \\
2MASXJ13003975+2755256	&		12.29$\pm$0.02	&11.18$\pm$0.02	&	3.37$\pm$0.09	&4.40$\pm$0.03	&	16.82$\pm$0.06	&16.20$\pm$0.02	&	2.240$\pm$0.009 \\
2MASXJ13002798+2757216	&		13.24$\pm$0.04	&12.26$\pm$0.05	&	2.29$\pm$0.06	&2.28$\pm$0.06	&	16.94$\pm$0.07	&15.86$\pm$0.08	&	2.121$\pm$0.011 \\
2MASXJ12595670+2755483	&		13.44$\pm$0.04	&12.49$\pm$0.05	&	4.16$\pm$0.32	&3.76$\pm$0.33	&	18.43$\pm$0.17	&17.16$\pm$0.20	&	2.001$\pm$0.016 \\
2MASXJ12594438+2754447	&		12.30$\pm$0.02	&11.25$\pm$0.02	&	3.44$\pm$0.03	&3.74$\pm$0.09	&	16.89$\pm$0.02	&15.93$\pm$0.06	&	2.207$\pm$0.009 \\
2MASXJ12594234+2755287	&		13.56$\pm$0.07	&12.53$\pm$0.05	&	1.07$\pm$0.21	&1.20$\pm$0.15	&	15.62$\pm$0.42	&14.74$\pm$0.27	&	2.238$\pm$0.011 \\
2MASXJ12594423+2757307	&		13.60$\pm$0.05	&12.68$\pm$0.07	&	2.99$\pm$0.12	&3.10$\pm$0.25	&	17.88$\pm$0.10	&16.95$\pm$0.19	&	1.819$\pm$0.025 \\
2MASXJ12593570+2757338	&		9.81$\pm$0.00	&8.98$\pm$0.00	&	26.13$\pm$0.03	&20.02$\pm$0.12	&	18.80$\pm$0.00	&17.30$\pm$0.01	&	2.426$\pm$0.009 \\
2MASXJ12590414+2757329	&		14.08$\pm$0.10	&12.98$\pm$0.11	&	1.74$\pm$0.31	&2.10$\pm$0.32	&	17.18$\pm$0.39	&16.40$\pm$0.35	&	2.112$\pm$0.014 \\
2MASXJ12565310+2755458	&		13.27$\pm$0.09	&12.26$\pm$0.04	&	1.21$\pm$0.08	&1.25$\pm$0.01	&	15.60$\pm$0.17	&14.59$\pm$0.05	&	2.272$\pm$0.009 \\
2MASXJ12562984+2756240	&		12.81$\pm$0.03	&11.70$\pm$0.03	&	2.48$\pm$0.68	&3.47$\pm$0.15	&	16.69$\pm$0.60	&16.23$\pm$0.10	&	2.266$\pm$0.009 \\
2MASXJ13012713+2759566	&		13.69$\pm$0.08	&12.60$\pm$0.05	&	0.87$\pm$0.15	&1.15$\pm$0.16	&	15.29$\pm$0.39	&14.70$\pm$0.31	&	2.198$\pm$0.010 \\
2MASXJ13005445+2800271	&		11.40$\pm$0.01	&10.46$\pm$0.01	&	7.55$\pm$0.03	&6.97$\pm$0.18	&	17.72$\pm$0.01	&16.54$\pm$0.06	&	2.371$\pm$0.009 \\
2MASXJ13003877+2800516	&		13.03$\pm$0.04	&11.97$\pm$0.03	&	3.86$\pm$0.08	&4.58$\pm$0.16	&	17.86$\pm$0.06	&17.07$\pm$0.08	&	2.026$\pm$0.014 \\
2MASXJ13000809+2758372	&		9.67$\pm$0.00	&8.51$\pm$0.00	&	14.79$\pm$0.15	&17.27$\pm$0.06	&	17.43$\pm$0.02	&16.52$\pm$0.01	&	2.570$\pm$0.008 \\
2MASXJ13000643+2800142	&		12.94$\pm$0.03	&12.02$\pm$0.04	&	3.25$\pm$0.06	&3.22$\pm$0.23	&	17.40$\pm$0.05	&16.37$\pm$0.16	&	2.081$\pm$0.013 \\
2MASXJ12594681+2758252	&		12.70$\pm$0.02	&11.53$\pm$0.02	&	3.55$\pm$0.09	&4.60$\pm$0.11	&	17.33$\pm$0.06	&16.60$\pm$0.05	&	2.116$\pm$0.012 \\
2MASXJ12593827+2759137	&		13.80$\pm$0.05	&13.11$\pm$0.10	&	3.80$\pm$0.27	&3.70$\pm$0.48	&	18.60$\pm$0.16	&17.77$\pm$0.30	&	1.909$\pm$0.022 \\
2MASXJ12592657+2759548	&		14.28$\pm$0.10	&13.06$\pm$0.09	&	1.77$\pm$0.23	&2.27$\pm$0.13	&	17.43$\pm$0.30	&16.66$\pm$0.16	&	1.904$\pm$0.018 \\
2MASXJ12592136+2758248	&		14.44$\pm$0.13	&13.48$\pm$0.14	&	2.26$\pm$0.22	&2.53$\pm$0.53	&	18.13$\pm$0.24	&17.34$\pm$0.47	&	1.713$\pm$0.050 \\
2MASXJ12590603+2759479	&		12.23$\pm$0.02	&11.30$\pm$0.02	&	4.68$\pm$0.25	&4.83$\pm$0.24	&	17.48$\pm$0.12	&16.52$\pm$0.11	&	2.179$\pm$0.010 \\
2MASXJ12583023+2800527	&		11.81$\pm$0.01	&11.14$\pm$0.02	&	5.87$\pm$0.19	&3.71$\pm$0.03	&	17.55$\pm$0.07	&15.80$\pm$0.03	&	2.282$\pm$0.009 \\
2MASXJ13025659+2804133	&		13.00$\pm$0.03	&12.22$\pm$0.05	&	4.59$\pm$0.25	&4.45$\pm$0.10	&	18.21$\pm$0.12	&17.26$\pm$0.07	&	1.753$\pm$0.030 \\
2MASXJ13024442+2802434	&		12.04$\pm$0.01	&10.98$\pm$0.02	&	5.22$\pm$0.08	&6.38$\pm$0.26	&	17.55$\pm$0.04	&16.84$\pm$0.09	&	2.254$\pm$0.010 \\
2MASXJ13004867+2805266	&		11.93$\pm$0.02	&10.89$\pm$0.01	&	3.28$\pm$0.04	&3.60$\pm$0.02	&	16.42$\pm$0.03	&15.48$\pm$0.02	&	2.321$\pm$0.008 \\
2MASXJ13002215+2802495	&		12.68$\pm$0.03	&11.54$\pm$0.03	&	3.48$\pm$0.09	&4.56$\pm$0.17	&	17.29$\pm$0.06	&16.62$\pm$0.08	&	2.087$\pm$0.011 \\
2MASXJ13001702+2803502	&		13.35$\pm$0.05	&12.35$\pm$0.04	&	2.68$\pm$0.26	&2.40$\pm$0.23	&	17.40$\pm$0.21	&16.09$\pm$0.21	&	2.070$\pm$0.013 \\
2MASXJ13001475+2802282	&		12.80$\pm$0.03	&11.80$\pm$0.03	&	1.91$\pm$0.05	&2.11$\pm$0.03	&	16.13$\pm$0.07	&15.27$\pm$0.04	&	2.206$\pm$0.010 \\
2MASXJ13001286+2804322	&		13.20$\pm$0.04	&12.35$\pm$0.05	&	2.15$\pm$0.10	&1.85$\pm$0.10	&	16.77$\pm$0.11	&15.48$\pm$0.12	&	2.075$\pm$0.011 \\
2MASXJ13000803+2804422	&		12.66$\pm$0.04	&11.57$\pm$0.02	&	1.83$\pm$0.04	&2.13$\pm$0.07	&	15.88$\pm$0.07	&15.02$\pm$0.08	&	2.263$\pm$0.008 \\
2MASXJ12595601+2802052	&		12.13$\pm$0.02	&11.04$\pm$0.02	&	3.99$\pm$0.13	&4.85$\pm$0.15	&	17.03$\pm$0.07	&16.25$\pm$0.07	&	2.206$\pm$0.010 \\
2MASXJ12593141+2802478	&		12.70$\pm$0.03	&11.86$\pm$0.04	&	3.29$\pm$0.13	&2.88$\pm$0.13	&	17.18$\pm$0.09	&15.98$\pm$0.10	&	2.108$\pm$0.011 \\
2MASXJ12591389+2804349	&		12.78$\pm$0.02	&12.11$\pm$0.05	&	6.60$\pm$0.11	&5.36$\pm$0.26	&	18.77$\pm$0.04	&17.55$\pm$0.12	&	2.150$\pm$0.014 \\
2MASXJ12564585+2803058	&		14.52$\pm$0.09	&13.51$\pm$0.12	&	2.70$\pm$0.18	&2.83$\pm$0.48	&	18.59$\pm$0.17	&17.59$\pm$0.39	&	1.669$\pm$0.048 \\
2MASXJ12563890+2804518	&		13.94$\pm$0.07	&13.17$\pm$0.11	&	4.35$\pm$0.32	&4.55$\pm$0.61	&	19.03$\pm$0.17	&18.25$\pm$0.31	&	1.625$\pm$0.062 \\
2MASXJ13014700+2805417	&		12.23$\pm$0.02	&11.11$\pm$0.02	&	2.67$\pm$0.02	&3.47$\pm$0.03	&	16.28$\pm$0.02	&15.66$\pm$0.02	&	2.257$\pm$0.008 \\
2MASXJ13004459+2806026	&		13.18$\pm$0.04	&12.46$\pm$0.05	&	4.14$\pm$0.28	&2.92$\pm$0.20	&	18.18$\pm$0.15	&16.61$\pm$0.16	&	1.891$\pm$0.017 \\
2MASXJ13003552+2808466	&		13.04$\pm$0.03	&11.96$\pm$0.03	&	3.91$\pm$0.22	&4.56$\pm$0.16	&	17.92$\pm$0.13	&17.11$\pm$0.08	&	1.885$\pm$0.020 \\
\hline
\end{tabular}
\end{center}
\end{table*}

\begin{table*} 
\begin{center}
Table \ref{Coma_JKs_phot} continued \ldots
\begin{tabular}{c@{\hskip 0.15in} r@{\hskip 0.15in} r@{\hskip 0.15in} r@{\hskip 0.15in} r@{\hskip 0.15in} c@{\hskip 0.15in} c@{\hskip 0.15in} c}  \\ \hline\hline
\multicolumn{1}{c}{Identification}  & \multicolumn{2}{c@{\hskip 0.15in}}{Total magnitude} &  \multicolumn{2}{c@{\hskip 0.15in}}{Effective radius} &  \multicolumn{2}{c@{\hskip 0.15in}}{Mean eff. surf. brightness} &  \multicolumn{1}{c}{$\log \sigma$} \\ 
 & \multicolumn{1}{c@{\hskip 0.15in}}{$J$}  & \multicolumn{1}{c@{\hskip 0.15in}}{$K_s$}  & \multicolumn{1}{c@{\hskip 0.15in}}{$J$} & \multicolumn{1}{c@{\hskip 0.15in}}{$K_s$} & \multicolumn{1}{c@{\hskip 0.15in}}{$J$} & \multicolumn{1}{c@{\hskip 0.15in}}{$K_s$}  &         \\ 
\multicolumn{1}{c@{\hskip 0.15in}}{(1)} & \multicolumn{1}{c@{\hskip 0.15in}}{(2)}  & \multicolumn{1}{c@{\hskip 0.15in}}{(3)}  & \multicolumn{1}{c@{\hskip 0.15in}}{(4)} & \multicolumn{1}{c@{\hskip 0.15in}}{(5)} & \multicolumn{1}{c@{\hskip 0.15in}}{(6)} & \multicolumn{1}{c@{\hskip 0.15in}}{(7)}   & \multicolumn{1}{c}{(8)}  \\ \hline
2MASXJ12595511+2807422	&		13.33$\pm$0.05	&12.25$\pm$0.04	&	2.32$\pm$0.27	&2.98$\pm$0.23	&	17.05$\pm$0.26	&16.42$\pm$0.17	&	2.096$\pm$0.014 \\
2MASXJ12590392+2807249	&		11.37$\pm$0.01	&10.35$\pm$0.01	&	4.77$\pm$0.13	&4.90$\pm$0.12	&	16.65$\pm$0.06	&15.59$\pm$0.05	&	2.422$\pm$0.008 \\
2MASXJ12585341+2807339	&		13.39$\pm$0.06	&12.52$\pm$0.05	&	2.34$\pm$0.11	&2.10$\pm$0.09	&	17.14$\pm$0.12	&15.95$\pm$0.11	&	2.062$\pm$0.013 \\
2MASXJ12583636+2806497	&		11.98$\pm$0.01	&11.25$\pm$0.02	&	5.19$\pm$0.09	&3.53$\pm$0.07	&	17.46$\pm$0.04	&15.80$\pm$0.04	&	2.239$\pm$0.009 \\
2MASXJ12574670+2808264	&		14.04$\pm$0.07	&13.41$\pm$0.14	&	4.16$\pm$0.45	&3.76$\pm$0.83	&	19.04$\pm$0.24	&18.11$\pm$0.50	&	1.607$\pm$0.066 \\
2MASXJ13021025+2811309	&		13.90$\pm$0.07	&13.05$\pm$0.11	&	3.95$\pm$0.39	&3.01$\pm$0.43	&	18.80$\pm$0.23	&17.30$\pm$0.33	&	1.834$\pm$0.025 \\
2MASXJ13012280+2811456	&		12.97$\pm$0.03	&12.12$\pm$0.04	&	4.00$\pm$0.53	&3.35$\pm$0.22	&	17.88$\pm$0.29	&16.55$\pm$0.15	&	2.109$\pm$0.013 \\
2MASXJ13001795+2812082	&		11.41$\pm$0.01	&10.12$\pm$0.01	&	4.26$\pm$0.02	&6.45$\pm$0.13	&	16.44$\pm$0.02	&15.95$\pm$0.04	&	2.369$\pm$0.008 \\
2MASXJ12592021+2811528	&		13.99$\pm$0.06	&12.88$\pm$0.05	&	2.32$\pm$0.10	&2.94$\pm$0.18	&	17.69$\pm$0.11	&16.97$\pm$0.14	&	1.949$\pm$0.022 \\
2MASXJ12581382+2810576	&		12.94$\pm$0.02	&12.13$\pm$0.05	&	6.46$\pm$0.11	&5.57$\pm$0.30	&	18.89$\pm$0.04	&17.67$\pm$0.12	&	1.869$\pm$0.019 \\
2MASXJ12574866+2810494	&		12.70$\pm$0.03	&11.69$\pm$0.03	&	2.72$\pm$0.11	&2.65$\pm$0.22	&	16.77$\pm$0.09	&15.62$\pm$0.18	&	2.159$\pm$0.011 \\
2MASXJ12572841+2810348	&		13.18$\pm$0.11	&11.69$\pm$0.03	&	2.71$\pm$0.20	&4.54$\pm$0.14	&	17.24$\pm$0.19	&16.76$\pm$0.07	&	2.021$\pm$0.015 \\
2MASXJ12563516+2816318	&		12.69$\pm$0.02	&12.03$\pm$0.04	&	4.67$\pm$0.24	&3.04$\pm$0.20	&	17.93$\pm$0.11	&16.26$\pm$0.15	&	2.090$\pm$0.013 \\
2MASXJ12592611+2817148	&		14.31$\pm$0.08	&13.58$\pm$0.12	&	2.02$\pm$0.12	&1.86$\pm$0.35	&	17.73$\pm$0.15	&16.72$\pm$0.42	&	1.695$\pm$0.039 \\
2MASXJ12584394+2816578	&		14.49$\pm$0.18	&13.78$\pm$0.17	&	2.26$\pm$0.38	&1.49$\pm$0.36	&	18.15$\pm$0.41	&16.45$\pm$0.55	&	1.629$\pm$0.046 \\
2MASXJ12582949+2818047	&		14.16$\pm$0.09	&13.29$\pm$0.12	&	2.76$\pm$0.25	&2.44$\pm$0.50	&	18.28$\pm$0.21	&17.07$\pm$0.46	&	1.844$\pm$0.027 \\
2MASXJ13024079+2822163	&		12.42$\pm$0.02	&11.45$\pm$0.03	&	5.40$\pm$0.15	&5.14$\pm$0.24	&	17.99$\pm$0.06	&16.81$\pm$0.11	&	1.941$\pm$0.020 \\
2MASXJ13021434+2821099	&		13.50$\pm$0.04	&12.62$\pm$0.08	&	5.46$\pm$0.19	&5.40$\pm$0.56	&	19.07$\pm$0.09	&18.05$\pm$0.24	&	1.855$\pm$0.032 \\
2MASXJ13020865+2823139	&		12.16$\pm$0.01	&11.01$\pm$0.02	&	4.61$\pm$0.11	&5.61$\pm$0.17	&	17.37$\pm$0.05	&16.56$\pm$0.07	&	2.213$\pm$0.020 \\
2MASXJ13010904+2821352	&		13.13$\pm$0.03	&12.51$\pm$0.05	&	4.88$\pm$0.39	&2.65$\pm$0.39	&	18.48$\pm$0.18	&16.44$\pm$0.33	&	1.879$\pm$0.018 \\
2MASXJ13005207+2821581	&		11.82$\pm$0.01	&10.80$\pm$0.01	&	3.89$\pm$0.10	&4.27$\pm$0.10	&	16.67$\pm$0.06	&15.75$\pm$0.05	&	2.240$\pm$0.010 \\
2MASXJ13004423+2820146	&		12.52$\pm$0.02	&11.78$\pm$0.03	&	3.41$\pm$0.03	&2.82$\pm$0.20	&	17.08$\pm$0.02	&15.82$\pm$0.16	&	2.098$\pm$0.010 \\
2MASXJ13003074+2820466	&		11.70$\pm$0.01	&10.78$\pm$0.01	&	6.63$\pm$0.08	&6.03$\pm$0.15	&	17.72$\pm$0.03	&16.53$\pm$0.06	&	2.205$\pm$0.009 \\
2MASXJ13023199+2826223	&		14.07$\pm$0.13	&13.44$\pm$0.16	&	5.11$\pm$0.84	&3.70$\pm$0.76	&	19.53$\pm$0.38	&18.12$\pm$0.48	&	1.776$\pm$0.064 \\
2MASXJ12575392+2829594	&		13.40$\pm$0.07	&12.46$\pm$0.06	&	1.15$\pm$0.28	&1.69$\pm$0.21	&	15.60$\pm$0.54	&15.42$\pm$0.28	&	2.121$\pm$0.012 \\
2MASXJ12593568+2833047	&		13.16$\pm$0.05	&12.25$\pm$0.04	&	2.07$\pm$0.42	&2.26$\pm$0.11	&	16.64$\pm$0.44	&15.82$\pm$0.11	&	2.113$\pm$0.012 \\
2MASXJ12565652+2837238	&		12.75$\pm$0.03	&11.97$\pm$0.03	&	4.05$\pm$0.19	&2.95$\pm$0.16	&	17.70$\pm$0.11	&16.14$\pm$0.13	&	2.036$\pm$0.013 \\
\hline
\end{tabular}
\end{center}
\end{table*}

\end{document}